\documentclass[aps,prb,showpacs,twocolumn,floats]{revtex4}
\usepackage{amssymb}

\usepackage{graphicx,psfig}
\usepackage{dcolumn}
\usepackage{bm}

\input{tcilatex}

\begin{document}

\title{Landau-Zener-Stueckelberg effect in a model of interacting tunneling
systems}
\author{D. A. Garanin}
\affiliation{ Institut f\"ur Physik,
Johannes-Gutenberg-Universit\"at,
 D-55099 Mainz, Germany}
\date{\today}

\begin{abstract}
The Landau-Zener-Stueckelberg (LZS) effect in a model system of
interacting tunneling particles is studied numerically and
analytically. Each of $N$ tunneling particles interacts with each
of the others with the same coupling $J.$ This problem maps onto
that of the LZS effect for a large spin $S=N/2.$ The mean-field
limit $N\rightarrow \infty $ corresponds to the classical limit
$S\rightarrow \infty $ for the effective spin. It is shown that
the ferromagnetic coupling $J>0$ tends to suppress the LZS
transitions. For $N\rightarrow \infty $ there is a critical value
of $J$ above which the staying probability $P$ does not go to zero
in the slow sweep limit, unlike the standard LZS effect. In the
same limit for $J>0$ LZS transitions are boosted and $P=0$ for a
set of finite values of the sweep rate. Various limiting cases
such as strong and weak interaction, slow and fast sweep are
considered analytically. It is shown that the mean-field approach
works well for arbitrary $N$ if the interaction $J$ is weak.
\end{abstract}


\maketitle

\section{Introduction}

The Landau-Zener-Stueckelberg (LZS)
problem\cite{lan32,zen32,stu32} of quantum-mechanical transitions
at an avoided level crossing induced by a linear-in-time energy
sweep is well known in many areas of physics, in particular, in
the physics of atomic and molecular collisions (see, e.g., Ref.\
\onlinecite{crohug77}). Recently the LZS\ effect was observed in
crystals of single-molecule magnets and it was used to extract
their tunneling level splittitng $\Delta
$.\cite{werses99,weretal00epl}

The specifics of the LZS effect in solid-state systems is the
macroscopically large number $N$ ot tunneling species such as spins of
magnetic molecules in the above experiments.\cite{werses99,weretal00epl}
These tunneling species are usually coupled with each other (mainly via the
dipole-dipole interaction in the case of magnetic molecules) and the
coupling energy $J$ can easily exceed the splitting $\Delta .$ In this
situation tunneling species (that can be described by pseudospins $S=1/2)$
do not tunnel independently, and one has to consider the Schr\"{o}dinger
equation for the whole system that is described by $2^{N}$ variables. The
latter is a new and tremendous problem since the whole Schr\"{o}dinger
equation becomes intractable even for a moderate number $N$ while analytical
treatment is difficult because the system is far from the equilibrium.

As a plausible first step towards the solution of the LZS problem for a
system of interacting particles one can consider a simplified model in which
each particle interacts with all $N-1$ other particles with the same
strength $J,$ as was done in Ref.\ \onlinecite{hamraemiysai00}. For this
model the Schr\"{o}dinger equation simplifies so that one has to solve only $%
N+1$ equations instead of the $2^{N}$ equations in the general case. In the
limit $N\rightarrow \infty $ the problem simplifies since the mean-filed
approximation (MFA) becomes exact. The latter allows one to gain insights
into the problem by analyzing its mean-field solution numerically as well as
analytically in different limiting cases. Because the \emph{molecular field}
exerted on a tunneling particle from the others depends nonlinearly on time
in the region of the level crossing, the problem cannot be linearized (c.f.
Ref. \onlinecite{hamraemiysai00}) and\ the results differ essentially from
the standard LZS solution. For instance, the ferromagnetic coupling, $J>0,$
makes the energy sweep faster, and thus the probability $P$ to stay on the
initial bare level increases. The nonlinear time dependence of the sweep
within the MFA has stimulated Ref.\ \onlinecite{garsch02prb}, where the
different kinds of sweep nonlinearities were investigated for a single
tunneling system. Certainly the MFA for systems of interacting tunneling
particles is a more complicated problem since in this case the form of the
sweep is unknown from the beginning and it should be found self-consistently.

It is very interesting to study the difference between the exact
quantum-mechanical and the mean-field solutions of the LZS problem for
finite $N.$ The original model maps of the model of a large spin $S=N/2$
with the uniaxial anisotropy $D\sim J$, transverse field $H_{x}\sim \Delta ,$
and the sweeped longitudinal field $H_{z}(t).$ That is, the mean-field limit
$N\rightarrow \infty $ of our model corresponds to the classical limit $%
S\rightarrow \infty $ for the effective large spin. In Ref.\ %
\onlinecite{hamraemiysai00} it was shown numerically that for the model with
$N=4$ and $J>0$ the exact and MFA results are qualitatively similar. It is
very interesting, however, to investigate the problem for larger $N$ and for
$J<0.$

Numerical and analytical study of these problems is the objective of this
paper, the rest of which is organized as follows. In Sec.\ \ref
{Sec-Hamiltonian} the Hamiltonian is written down, the Schr\"{o}dinger
equation is simplified making use of the symmetry of the interaction, the
problem is mapped onto that of a large spin and its general properties are
studied. In Sec.\ \ref{Sec-separated} the case of well-separated resonances,
$\left| J\right| \gg \Delta $ where the problem can be reduced to that of
successive standard LZS transitions is investigated. In Sec.\ \ref
{Sec-fast-sweep} the opposite case $J\ll \Delta ,$ where the problem can be
solved perturbatively in $J,$ is considered. Final results are worked out in
the cases of fast sweep and slow sweep. In Sec.\ \ref{Sec-mfa} general
properties of the mean-field solution of the LZS problem with interaction
are investigated, in particular, with the help of the mapping onto the
classical-spin problem. Numerical solution shows, in particular, that for $%
J<0$ complete LZS transition, $P=0,$ is achieved at some values of the sweep
rate. Analytical treatment of the slow-sweep limit within the MFA is
provided in Sec.\ \ref{Sec-slow-MFA}.

\section{The Hamiltonian}

\label{Sec-Hamiltonian}

We consider the model of $N$ double-level tunneling systems described by
pseudospins 1/2 and interacting each with each with an equal strength
\begin{equation}
\widehat{H}=-\frac{1}{2}\sum_{i=1}^{N}\left[ W(t)\sigma _{iz}+\Delta \sigma
_{ix}\right] -\frac{J}{2}\sum_{i\neq j=1}^{N}\sigma _{iz}\sigma _{jz}.
\label{Ham}
\end{equation}
Here $\mathbf{\sigma }_{i}$ are Pauli matrices, $W(t)$ is the energy sweep
that is taken to be linear in time
\begin{equation}
W(t)\equiv E_{-1}(t)-E_{1}(t)=vt,  \label{WDef}
\end{equation}
$\Delta $ is the level splitting at resonance in the absence of interaction (%
$t=0$ and $J=0)$, whereas $J$ is the interaction constant. The case $J>0$
corresponds to the ferromagnetic (FM) coupling, whereas that of $J<0$
corresponds to the antiferromagnetic frustrating (AFMF) coupling. In the
latter case, the ground state of the system with $W=\Delta =0$ is a highly
degenerate state with the minimal total spin. If $J=0,$ the problem
simplifies to the well known LZS problem for individual tunneling systems.
If these system at $t\rightarrow -\infty $ are in the bare ground state $%
\psi _{-1}\equiv |\downarrow \rangle $ before crossing the resonance, the
probability to stay in this state after crossing the resonance at $%
t\rightarrow \infty $ \ is given by \cite{zen32,stu32,akusch92,dobzve97}
\begin{equation}
P(\infty )\equiv P=e^{-\varepsilon },\qquad \varepsilon \equiv \frac{\pi
\Delta ^{2}}{2\hbar v}  \label{PLZ}
\end{equation}
(Probabilities without the argument are shortcuts for the final-state
probabilities at $t=\infty $ throughout the paper). For the problem with
interaction, the wave function of the system can be written as the expansion
over the direct-product states
\begin{eqnarray}
\Psi (t) &=&\sum_{m_{1},\ldots ,m_{N}=-1,1}C_{m_{1},\ldots ,m_{N}}(t)\psi
_{m_{1}}\otimes \ldots \otimes \psi _{m_{N}}  \nonumber \\
\psi _{-1} &=&\left(
\begin{array}{c}
0 \\
1
\end{array}
\right) =|\downarrow \rangle ,\qquad \psi _{1}=\left(
\begin{array}{c}
1 \\
0
\end{array}
\right) =|\uparrow \rangle .  \label{PsiDef}
\end{eqnarray}
The initial condition for $\Psi (t)$ is $C_{-1,\ldots ,-1}(-\infty )=1$
whereas all other coefficients are zero, i.e., the system starts in the
state with all spins down. With time the state of the system becomes a
superposition of all possible basis states in Eq.\ (\ref{PsiDef}). The
one-particle probability to remain in the initial state $-1$ for our $N$%
-particle system is given by
\begin{equation}
P_{N}(t)=\sum_{m_{2},\ldots ,m_{N}=-1,1}\left| C_{-1,m_{2},\ldots
,m_{N}}(t)\right| ^{2}  \label{PtDef}
\end{equation}
and it starts from $P_{N}(-\infty )=1.$

The solution of the LZS problem for this model is simplified by the fact
that the coefficients $C_{m_{1},\ldots ,m_{N}}(t)$ depend only on the number
$k$ of spins up while they are independent on the choice of these spins.
Thus one can label the states by the index $k$ only that results in the
Schr\"{o}dinger equation
\begin{equation}
i\hbar \dot{C}_{k}=E_{k}(t)C_{k}-\frac{k\Delta }{2}C_{k-1}-\frac{(N-k)\Delta
}{2}C_{k+1}  \label{SchrEq}
\end{equation}
for $k=0,\ldots ,N$ and with the bare energies
\begin{equation}
E_{k}(t)=\left( \frac{N}{2}-k\right) W(t)+2Jk(N-k)+\mathrm{const}.
\label{Enk}
\end{equation}
Each $k$-state is realized by $N!/\left[ (N-k)!k!\right] $ different choices
of the indices $m_{1},\ldots ,m_{N}$ in Eq.\ (\ref{PsiDef}). For $J<0,$ the
ground state with $W=\Delta =0$\ corresponds to $k=N/2$ and is highly
degenerate. In Eq.\ (\ref{PtDef}) $m_{1}$ is fixed to $-1,$ and the
remaining indices $m_{2},\ldots ,m_{N}$ can also be parametrized by $k.$ In
the corresponding number of realizations, one should use $N-1$ instead of $N$
that leads to
\begin{equation}
P_{N}(t)=\sum_{k=0}^{N-1}\frac{(N-1)!}{(N-1-k)!k!}\left| C_{k}(t)\right|
^{2}.  \label{PtDefk}
\end{equation}

It is convenient to introduce the coefficients
\begin{equation}
c_{k}\equiv \sqrt{\frac{N!}{(N-k)!k!}}C_{k}  \label{ckDef}
\end{equation}
that satisfy the normalization condition $1=\sum_{k=0}^{N}\left|
c_{k}\right| ^{2}$ and the Schr\"{o}dinger equation
\begin{equation}
i\hbar \dot{c}_{k}=E_{k}(t)c_{k}-\frac{\Delta }{2}l_{k-1,k}c_{k-1}-\frac{%
\Delta }{2}l_{k,k+1}c_{k+1}  \label{SchrEqc}
\end{equation}
with $l_{k,k+1}\equiv \sqrt{(N-k)(k+1)}.$ Then Eq.\ (\ref{PtDefk})
simplifies to
\begin{equation}
P_{N}(t)=\sum_{k=0}^{N}\left( 1-\frac{k}{N}\right) p_{k}(t),\qquad
p_{k}(t)\equiv \left| c_{k}(t)\right| ^{2}.  \label{PtFinal}
\end{equation}
Eq.\ (\ref{SchrEqc}) maps on the corresponding Schr\"{o}dinger equation for
a large pseudospin $S=N/2$ with the Hamiltonian
\begin{equation}
\widehat{H}_{S}=-H_{z}(t)S_{z}-H_{x}S_{x}-DS_{z}^{2}+\mathrm{const,}
\label{HamS}
\end{equation}
where
\begin{equation}
H_{z}(t)=W(t),\qquad H_{x}=\Delta ,\qquad D=2J  \label{ParIdent}
\end{equation}
(we set $g\mu _{B}=1$ so that the ``magnetic field'' is energy dimensional).
Note that the excitation number $k$ is related to the eigenvalue $m$ of $%
S_{z}$ by $k=N/2+m.$ Thus $P(t)$ maps onto
\begin{equation}
P_{N}(t)=\frac{1}{2}\left( 1-\frac{\langle S_{z}\rangle _{t}}{S}\right) .
\label{PtMapping}
\end{equation}

\begin{figure}[t]
\unitlength1cm
\begin{picture}(11,6)
\centerline{\psfig{file=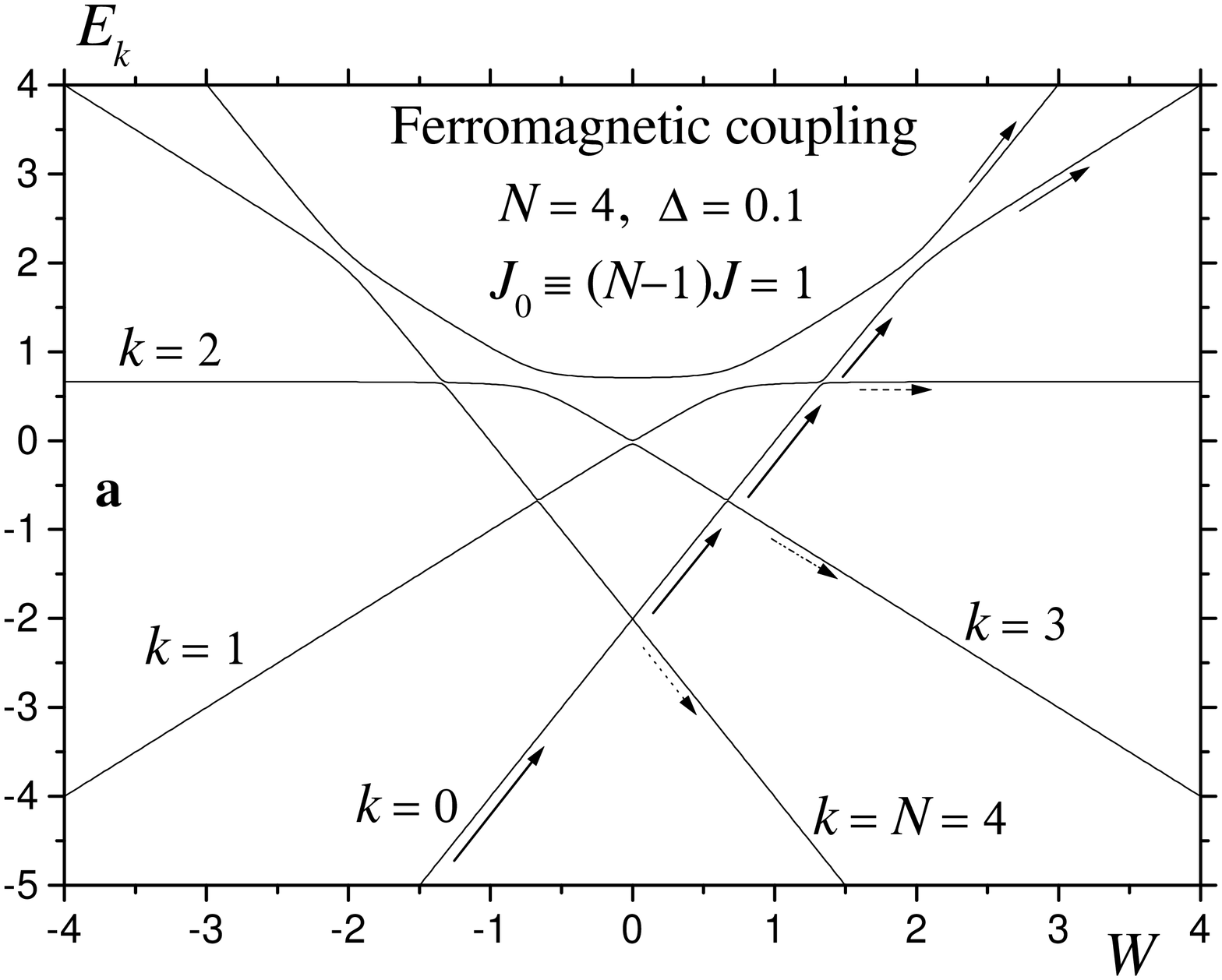,angle=-90,width=9cm}}
\end{picture}
\begin{picture}(11,6)
\centerline{\psfig{file=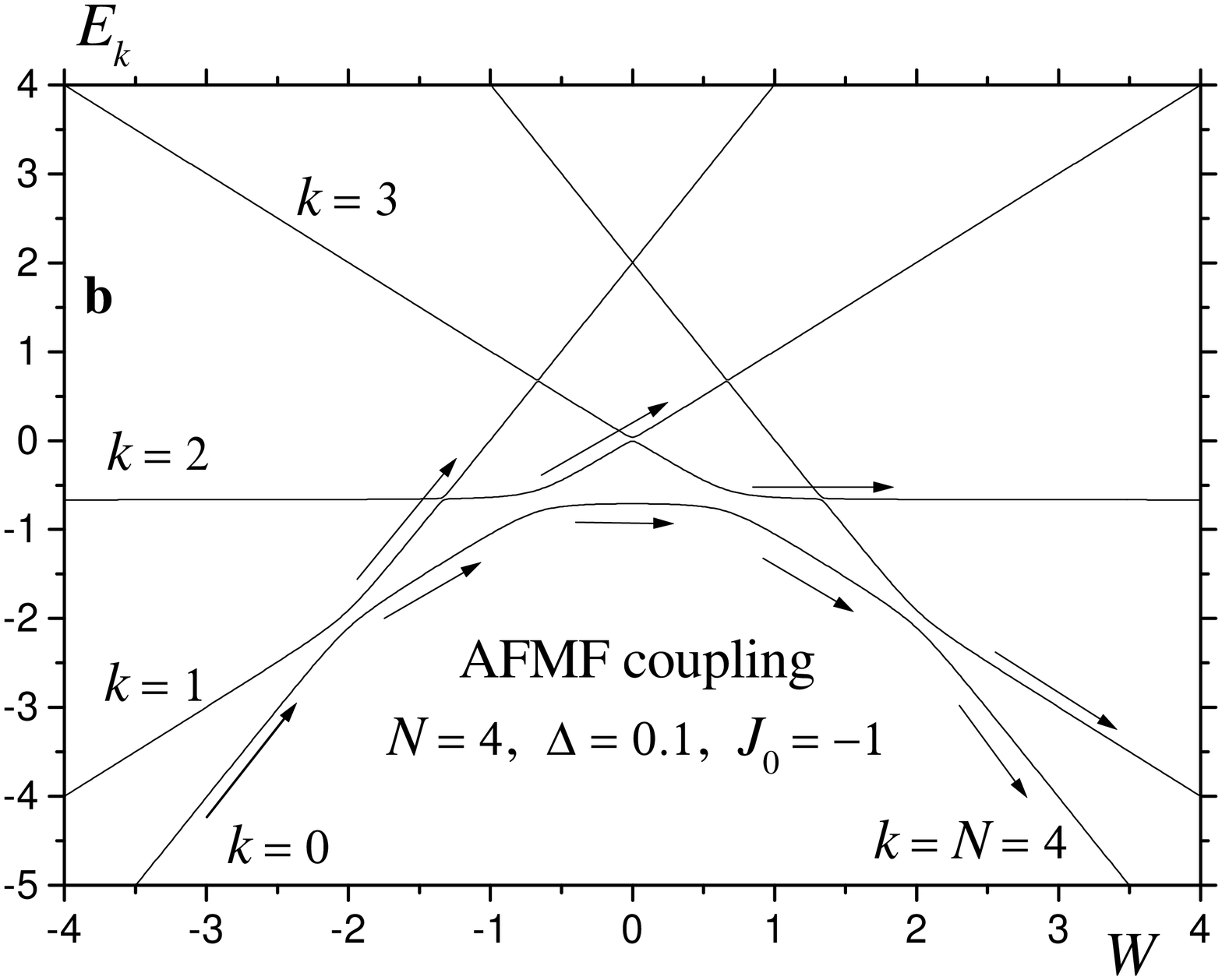,angle=-90,width=9cm}}
\end{picture}
\caption{ \label{Fig-lzn-En}
$a$ -- Exchange-split resonances for a ferromagnetically coupled spin cluster for $|J|\gtrsim \Delta$.
Transitions shown by the dashed and dotted arrows arise in higher orders in $\Delta/J$ and are much weaker.
$b$ -- Same for  an AFMF coupled spin cluster.
}
\end{figure}

The large-spin model defined by Eq.\ (\ref{HamS}) is well known from the
physics of the single-molecule magnet Mn$_{12}.$ For small $\Delta $ the
energy levels of the system as function of $W$ are nearly straight lines
with avoided crossings (see Fig.\ \ref{Fig-lzn-En}). Splittings at avoided
crossings are or order $\Delta (\Delta /J)^{n},$ where $n$ are appropriate
powers.\cite{gar91jpa}

%TCIMACRO{
%\TeXButton{Fig-lzn-En}{\begin{figure}[t]
%\unitlength1cm
%\begin{picture}(11,6)
%\centerline{\psfig{file=Fig-lzn-En-F.eps,angle=-90,width=9cm}}
%\end{picture}
%\begin{picture}(11,6)
%\centerline{\psfig{file=Fig-lzn-En-AF.eps,angle=-90,width=9cm}}
%\end{picture}
%\caption{ \label{Fig-lzn-En}
%$a$ -- Exchange-split resonances for a ferromagnetically coupled spin cluster for $|J|\gtrsim \Delta$.
%Transitions shown by the dashed and dotted arrows arise in higher orders in $\Delta/J$ and are much weaker.
%$b$ -- Same for  an AFMF coupled spin cluster.
%}
%\end{figure}%
%}}%
%BeginExpansion
%
%EndExpansion

%TCIMACRO{
%\TeXButton{Fig-lzn-P3Ft}{\begin{figure}[t]
%\unitlength1cm
%\begin{picture}(11,6)
%\centerline{\psfig{file=Fig-lzn-P3Ft.eps,angle=-90,width=9cm}}
%\end{picture}
%\caption{ \label{Fig-lzn-P3Ft}
%Time dependence of the one-particle staying probability $P_N$ as function of the energy sweep $W(t)$
%for a system of $N=3$ ferromagnetically coupled particles.
%At intermediate sweep rate, $\varepsilon=1$ only the first-order LZS transition at $W=2J_0$ (the $\Delta$ transition) is seen.
%For rather slow sweep rates (here  $\varepsilon=10$),  higher-order transitions (here the  $\Delta^2/J$ transition at $W=0$) are switching on.
%}
%\end{figure}%
%}}%
%BeginExpansion
\begin{figure}[t]
\unitlength1cm
\begin{picture}(11,6)
\centerline{\psfig{file=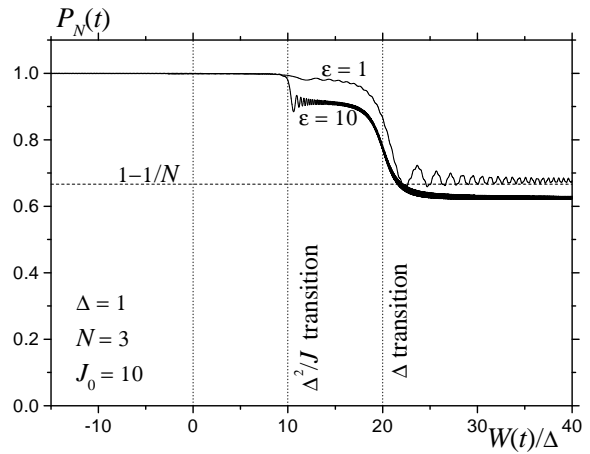,angle=-90,width=9cm}}
\end{picture}
\caption{ \label{Fig-lzn-P3Ft}
Time dependence of the one-particle staying probability $P_N$ as function of the energy sweep $W(t)$
for a system of $N=3$ ferromagnetically coupled particles.
At intermediate sweep rate, $\varepsilon=1$ only the first-order LZS transition at $W=2J_0$ (the $\Delta$ transition) is seen.
For rather slow sweep rates (here  $\varepsilon=10$),  higher-order transitions (here the  $\Delta^2/J$ transition at $W=0$) are switching on.
}
\end{figure}%
%
%EndExpansion

Our model described by Eqs.\ (\ref{Ham}) or (\ref{HamS}), although not
completely realistic, allows one to trace down the influence of the
interaction on the transition probabilities and to obtain a number of
interesting analytical results. We will see that, in particular, suppression
of transitions for the ferromagnetic interactions ($J>0)$ can be easily
understood. An advantage of this simplified model is the possibility to
solve the problem numerically up to much larger system sizes, compared to
models with realistic interactions. For the latter, the number of different
coefficients in Eq.\ (\ref{PsiDef}) and thus of differential equations to
solve is $2^{N},$ whereas in our case there are only $N+1$ equations.

For numerical calculations we use Wolfram Mathematica 4.0 that employs, in
particular, a very accurate differential equation solver needed for handling
strongly oscillating solutions over large time intervals. The results of
solving Eq.\ (\ref{SchrEqc}) are shown in Fig.\ \ref{Fig-lzn-P3Ft} for a
cluster of three ferromagnetically coupled particles.

\section{Well-separated resonances}

\label{Sec-separated}

The interaction can profoundly influence the LZS effect if it is strong
enough, \TEXTsymbol{\vert}$J|\gtrsim \Delta $. The general tendency can be
already seen from the well-known mean-field argument. If one of the
tunneling particles flips to another bare state during the resonance
crossing, then the total field (the external sweep field plus the molecular
field) on all other particles jumps. For the ferromagnetic coupling, $J>0,$
the jump of the total field is positive, other particles are brought past
the resonance and lose their chance to flip, and thus transition probability
is suppressed. For the antiferromagnetic frustrating coupling, the jump of
the total field is negative and other particles are being set back before
the resonance and are getting one more chance to cross the resonance at some
larger field value and flip; Thus the transition probability for the system
should increase.

The influence of the interaction can be readily illustrated within a
rigorous many-body quantum language for our model if one considers the
energy levels of the system shown in Fig.\ \ref{Fig-lzn-En}. One can see
that instead of a single resonance at $W=0$ for individual or noninteracting
particles there is an interaction-split resonance that consists of many
avoided line crossings. These avoided line crossings are well separated from
each other on the plot if \TEXTsymbol{\vert}$J|\gg \Delta .$ The question
whether these well-separated crossings can be considered as a succession of
independent LZS transitions (i.e., whether they are \emph{dynamically} well
separated) depends on the sweep rate. Whereas for the slow sweep, $%
\varepsilon \gtrsim 1$ [see Eq.\ (\ref{PLZ})] the condition \TEXTsymbol{\vert%
}$J|\gg \Delta $ is sufficient, for the fast sweep, $\varepsilon \lesssim 1,$
a more stringent condition is required that follows from Eq.\ (16) of Ref.\ %
\onlinecite{garsch02prb}. The resulting combined condition for the
dynamically well-separated resonances thus would be
\begin{equation}
|J|\gg \left\{
\begin{array}{cc}
\Delta , & \varepsilon \gtrsim 1 \\
\sqrt{\hbar v}\sim \Delta /\varepsilon , & \varepsilon \lesssim 1.
\end{array}
\right.  \label{SepResCond}
\end{equation}
This is, however, only an \emph{apriori} estimation considering two
resonances. For $N\gg 1$ there are many resonances, and the deviations from
the single-resonance picture can accumulate with the increase of $N.$ One
can do an \emph{aposteriori} estimation in the slow-sweep limit where the
result for the well-separated resonances $P\rightarrow 1-1/N$ for $%
\varepsilon \rightarrow \infty $ following from Eq.\ (\ref{PNF}) as well as
the mean-field result for $N\gg 1$ and non-separated resonances [the first
line of Eq.\ (\ref{PhxLims})] are available, in both cases $1-P\ll 1$. The
crossover between these results should take place at \TEXTsymbol{\vert}$%
J|\sim $ $\Delta N^{1/2},$ thus the first line of Eq.\ (\ref{SepResCond})
should be replaced by
\begin{equation}
|J|\gg \Delta N^{1/2},\qquad \varepsilon \gtrsim 1.  \label{SepResCond2}
\end{equation}
The second line of Eq.\ (\ref{SepResCond}) could be also modified by $N$ but
it is difficult to derive an exact condition.

A perturbative analysis of the stationary Schr\"{o}dinger equation for our
model shows that tunnel level splitting of the bare levels with different
values of $k$ has the form\cite{gar91jpa}
\begin{equation}
\delta E_{k_{1},k_{2}}\sim \Delta \left( \frac{\Delta }{J}\right)
^{|k_{1}-k_{2}|-1}.  \label{SplittDiffOrd}
\end{equation}
Thus for \TEXTsymbol{\vert}$J|\gg \Delta $ only the first-order or direct
transitions with $|k_{1}-k_{2}|=1$ should be taken into account whereas the
higher-order transitions are weak (see Fig.\ \ref{Fig-lzn-En}).

Now it becomes clear that for the ferromagnetic coupling and well-separated
resonances only the strong transition between the initial level $k=0$ and
the level $k=1$ happens. That is, in another language, only one particle of $%
N$ has a chance to tunnel, and the tunneling probability for the whole
system is strongly reduced. For a quantitative analysis one can neglect all $%
c_{k}$ with $k>1$ in Eq.\ (\ref{SchrEqc}) that yields the system of
equations
\begin{eqnarray}
i\hbar \dot{c}_{0} &=&E_{0}c_{0}-\frac{\Delta }{2}\sqrt{N}c_{1}  \nonumber \\
i\hbar \dot{c}_{1} &=&E_{1}c_{1}-\frac{\Delta }{2}\sqrt{N}c_{0}
\label{SchrEqcF}
\end{eqnarray}
that maps on the standard LZS problem with $\Delta \Rightarrow \Delta
_{N}\equiv \Delta \sqrt{N}$ and with the resonance at $W=2(N-1)J$ instead of
$W=0.$ The final-state probabilities for Eq.\ (\ref{SchrEqcF}) are according
to Eq.\ (\ref{PLZ})
\begin{equation}
p_{0}=P^{N}=e^{-N\varepsilon },\qquad p_{1}=1-P^{N}.  \label{p0p1F}
\end{equation}
Then with the help of Eq.\ (\ref{PtFinal}) the one-particle staying
probability for $J>0$ can be cast into the form
\begin{equation}
P_{N}=1-\frac{1}{N}\left( 1-P^{N}\right)  \label{PNF}
\end{equation}
and it varies between $1$ in the fast-sweep limit and $1-1/N$ in the
slow-sweep limit. Thus suppression of transitions by the ferromagnetic
coupling is extremely strong for a large number of particles $N.$ Expansion
of the transition probability of Eq.\ (\ref{PNF}) in the fast-sweep limit
reads
\begin{equation}
1-P_{N}\cong \varepsilon -\frac{N}{2}\varepsilon ^{2}+O(\varepsilon
^{3}),\qquad (\varepsilon \ll 1/N).  \label{PNFfast}
\end{equation}
Note that the first term of this expansion is insensitive to the interaction
(in this context to the number of interacting particles) and is the same as
for the standard LZS problem, $1-P=1-e^{-\varepsilon }\cong \varepsilon
-\varepsilon ^{2}/2!+\varepsilon ^{3}/3!-\ldots $

%TCIMACRO{
%\TeXButton{Fig-lzn-separatedresonances}{\begin{figure}[t]
%\unitlength1cm
%\begin{picture}(11,6)
%\centerline{\psfig{file=Fig-lzn-separatedresonances.eps,angle=-90,width=9cm}}
%\end{picture}
%\caption{ \label{Fig-lzn-separatedresonances}
%One-particle staying probability $P_N$ vs the sweep-rate parameter $\varepsilon$ for systems with FM and AFMF interaction in the case of well separated resonances.
%}
%\end{figure}%
%}}%
%BeginExpansion
\begin{figure}[t]
\unitlength1cm
\begin{picture}(11,6)
\centerline{\psfig{file=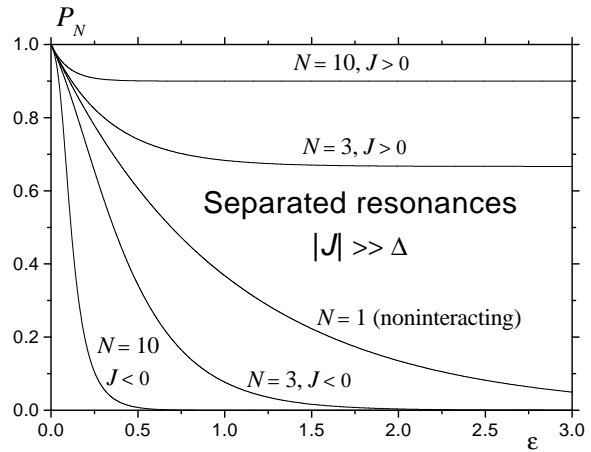,angle=-90,width=9cm}}
\end{picture}
\caption{ \label{Fig-lzn-separatedresonances}
One-particle staying probability $P_N$ vs the sweep-rate parameter $\varepsilon$ for systems with FM and AFMF interaction in the case of well separated resonances.
}
\end{figure}%
%
%EndExpansion

The AFMF coupling $J<0$ favors transitions, as can be seen from Fig.\ \ref
{Fig-lzn-En}b. The original resonance is splitted by the interaction into $N$
resonances filling equidistantly the range $-2(N-1)J\leq W\leq 2(N-1)J.$ One
can see that in the limit of slow sweep the system remains on the lowest
adiabatic energy level thus $P_{N}\rightarrow 0,$ in contrast to the
ferromagnetic case. For well-separated resonances the problem described by
Eq.\ (\ref{SchrEqc}) splits into independent standard LZS problems for the
resonances between the levels $k$ and $k+1$ that are described by an
effective splitting $\Delta _{N,k}=\Delta l_{k,k+1}$ [cf. Eq.\ (\ref
{SchrEqcF})]. It is convenient to designate the probability to stay in state
$k$ after crossing with state $k-1$ but before crossing with state $k+1$
(see Fig.\ \ref{Fig-lzn-En}b) as $p_{k}($\textrm{mid}$)$. Then solutions of
the LZS problems for individual crossings along with conditions of the
probability conservation can be written as
\begin{eqnarray}
p_{k}(\infty ) &=&P^{(N-k)(k+1)}p_{k}(\text{\textrm{mid}})  \nonumber \\
p_{k}(\text{\textrm{mid}}) &=&p_{k-1}(\text{\textrm{mid}})-p_{k-1}(\infty )
\nonumber \\
p_{k-1}(\infty ) &=&P^{(N-k+1)k}p_{k-1}(\text{\textrm{mid}}),  \label{EqsAF}
\end{eqnarray}
etc. These can be combined into the recurrence relation
\begin{eqnarray}
p_{k}(\infty ) &=&P^{(N-k)(k+1)}\left( P^{-(N-k+1)k}-1\right) p_{k-1}(\infty
)  \nonumber \\
&=&P^{N-2k}\left( 1-P^{(N-k+1)k}\right) p_{k-1}(\infty )  \label{RecurrAF}
\end{eqnarray}
that has to be iterated with the initial condition $p_{0}(\infty )=P^{N}.$
In the slow-sweep limit $\varepsilon \gg 1/N,$ one has $P^{N}\ll 1,$ thus
one can drop the factor $\left( 1-P^{(N-k+1)k}\right) \cong 1$ in the second
line of Eq.\ (\ref{RecurrAF}) that after iteration results in $p_{k}\cong
P^{(N-k)(k+1)}.$ In this case, in Eq.\ (\ref{PtFinal}) $p_{0}\cong
p_{N-1}\cong P^{N}$ are dominant for large $N$ whereas all other summands
are much smaller. This yields
\begin{equation}
P_{N}\cong \left( 1+\frac{1}{N}\right) P^{N}\ll 1,\qquad \left( \varepsilon
\gg 1/N\right)  \label{PNAFslow}
\end{equation}
for the antiferromagnetic coupling. In the fast-sweep limit the solution of
Eq.\ (\ref{RecurrAF}) can be expanded in powers of $\varepsilon ,$ after
which Eq.\ (\ref{PtFinal}) yields for the transition probability
\begin{equation}
1-P_{N}\cong \varepsilon +\left( \frac{3N}{2}-2\right) \varepsilon
^{2}+O(\varepsilon ^{3}),\quad (\varepsilon \ll 1/N).  \label{PNAFfast}
\end{equation}
Again, the first term of this expansion is the same as in the standard LZS
problem. For intermediate sweep rates one can iterate Eq.\ (\ref{RecurrAF})
numerically and substitute the solution for $p_{k}$ into Eq.\ (\ref{PtFinal}%
). The resulting curves $P_{N}$ vs $\varepsilon $ are shown in Fig.\ \ref
{Fig-lzn-separatedresonances} along with those for the ferromagnetic
coupling, Eq.\ (\ref{PNF}).

\begin{figure}[t]
\unitlength1cm
\begin{picture}(11,6)
\centerline{\psfig{file=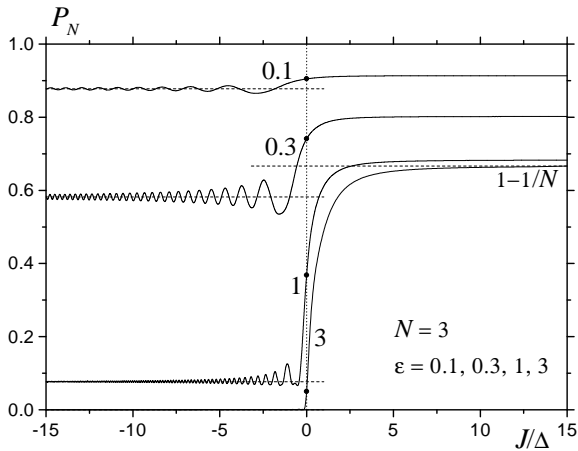,angle=-90,width=9cm}}
\end{picture}
\caption{ \label{Fig-lzn-PvsJ}
Numerically computed staying probability $P_{N}$ vs $J/\Delta $ for $N=3$ and different sweep rate parameters $\varepsilon $.
The horizontal dashed lines on the left side of the plot are asymptotes corresponding to well-separated resonances.
}
\end{figure}

To conclude this section, we show the numerically computed dependences of $%
P_{N}$ on $J/\Delta $ for $N=3$ and different $\varepsilon $ in Fig.\ \ref
{Fig-lzn-PvsJ}. The case of well-separated resonances that was considered in
this section corresponds to the plateaus on the left and on the right sides
of the plot.

%TCIMACRO{
%\TeXButton{Fig-lzn-PvsJ}{\begin{figure}[t]
%\unitlength1cm
%\begin{picture}(11,6)
%\centerline{\psfig{file=Fig-lzn-PvsJ.eps,angle=-90,width=9cm}}
%\end{picture}
%\caption{ \label{Fig-lzn-PvsJ}
%Numerically computed staying probability $P_{N}$ vs $J/\Delta $ for $N=3$ and different sweep rate parameters $\varepsilon $.
%The horizontal dashed lines on the left side of the plot are asymptotes corresponding to well-separated resonances.
%}
%\end{figure}%
%}}%
%BeginExpansion
%
%EndExpansion

\section{Fast sweep and weak coupling}

\label{Sec-fast-sweep}

In the preceding section we have considered the strong-coupling limit
\TEXTsymbol{\vert}$J|\gg \max \left( \Delta N^{1/2},\Delta /\sqrt{%
\varepsilon }\right) $ in which individual resonances are well separated
from each other and one thus deals with successive standard LZS transitions.
The opposite limiting case is that of the weak coupling $J.$ This case can
be solved by a perturbative expansion in $J$ with the standard LZS solution
as the zeroth-order appriximation. The situation further simplifies in the
fast-sweep limit, where the transition probability is small and the
coefficients $c_{k}$ in Eq.\ (\ref{SchrEqc}) decrease with $k$ as powers of $%
\varepsilon \ll 1.$ In particular, $c_{0}$ contains terms of orders $%
\varepsilon ^{0},\varepsilon ^{1},$ and $\varepsilon ^{2},$ the coefficient $%
c_{1}$ contains $\varepsilon ^{1/2}$ and $\varepsilon ^{3/2},$ the
coefficient $c_{2}$ starts with $\varepsilon ,$ etc. Knowing these
contributions into $c_{0},$ $c_{1},$ and $c_{2}$ is sufficient to set up the
expansion of $P_{N}$ up to the first nontrivial order $\varepsilon ^{2}.$ A
straighforward but cumbersome calculation yields
\begin{eqnarray}
1-P_{N} &\cong &\varepsilon -\left( \frac{1}{2}+\frac{4J_{0}}{\sqrt{2\pi
\hbar v}}\right) \varepsilon ^{2}+O(\varepsilon ^{3})  \nonumber \\
&=&\varepsilon -\frac{1}{2}\varepsilon ^{2}-\frac{4J_{0}}{\Delta }%
\varepsilon ^{5/2}+\ldots  \label{PNSmallJfast}
\end{eqnarray}
where we have defined
\begin{equation}
J_{0}\equiv (N-1)J.  \label{J0Def}
\end{equation}
One can see, again that ferromagnetic interaction suppresses transitions
whereas the AFMF interaction facilitates transitions, and that the effect of
interaction is increased by the number particles in the system. In fact,
however, in our model the interaction should scale, on physical grounds,
with the system's size, i.e., $J_{0}\equiv $ $(N-1)J$ should be size
independent. Note that in the fast-sweep limit one could do the expansion in
$\varepsilon $ for arbitrary $J.$ This leads, however, to cumbersome
multiple integrals. The strong-coupling limit for the fast sweep $%
\varepsilon \ll 1$ has been considered above, and the corresponding
counterparts of Eq.\ (\ref{PNSmallJfast}) for ferro-and antiferromagnetic
coupling are Eqs.\ (\ref{PNFfast}) and (\ref{PNAFfast}).

It should be noted that although Eq.\ (\ref{PNSmallJfast}) is valid for $%
\varepsilon \ll 1,$ the method of its derivation above is only justified for
$\varepsilon \ll 1/N,$ i.e. for much faster sweep rates, if $N\gg 1.$
Indeed, in the course of the derivation the terms of orders $\left(
\varepsilon N\right) ^{1/2},$ $\varepsilon N,$ etc., appear that are
required to be small. Only at the very end the leading $N$ contributions
cancel each other that leads to Eq.\ (\ref{PNSmallJfast}) that fortunately
has a larger applicability range. In fact, for large systems with weak
interaction the most populated final states are
\begin{equation}
k_{\max }=N(1-P)\cong N\varepsilon ,  \label{kmaxDef}
\end{equation}
so that keeping only the states with $k=0,1,2$ above was not justified for $%
N\varepsilon \gtrsim 1.$ Eq.\ (\ref{kmaxDef}) can be easily derived for an
assembly of noninteracting tunneling species. To this end, we use the
coefficients $C_{k}$ of Eq.\ (\ref{SchrEq}) and represent them in the
factorized form
\begin{equation}
C_{k}=a_{-1}^{N-k}a_{1}^{k},  \label{CFactorized}
\end{equation}
where $a_{-1}$ and $a_{1}$ are the coefficients of the wave function of the
one-particle problem, $\psi =a_{-1}\psi _{-1}+a_{1}\psi _{1}.$ This yields
\begin{equation}
|C_{k}|^{2}=|a_{-1}|^{2\left( N-k\right) }|a_{1}|^{2k}=P^{N-k}(1-P)^{k}
\label{C2Factorized}
\end{equation}
that can be plugged into Eq.\ (\ref{PtDefk}) to give
\begin{equation}
P_{N}=\sum_{k=0}^{N-1}\frac{(N-1)!}{(N-1-k)!k!}P^{N-k}(1-P)^{k}.
\label{PNNoninteracting}
\end{equation}
The summand in this formula has its maximum at $k=k_{\max }$ given by Eq.\ (%
\ref{kmaxDef}) and it becomes a narrow Gaussian packet around $k_{\max }$\
for $N\gg 1.$

A more rigorous method of handling the fast-sweep and weak coupling limits
uses the one-particle density matrix that is defined by
\begin{equation}
\rho _{-1,-1}=P_{N},\qquad \rho _{1,1}=1-\rho _{-1,-1}=1-P_{N}
\label{rhodiagDef}
\end{equation}
with $P_{N}$ of Eq.\ (\ref{PtDefk}) and
\begin{eqnarray}
\rho _{1,-1} &=&\sum_{m_{2},\ldots ,m_{N}=-1,1}C_{1,m_{2},\ldots
,m_{N}}^{\ast }C_{-1,m_{2},\ldots ,m_{N}}  \nonumber \\
&=&\sum_{k=0}^{N-1}\frac{(N-1)!}{(N-1-k)!k!}C_{k+1}^{\ast }C_{k}  \nonumber
\\
\rho _{-1,1} &=&\rho _{1,-1}^{\ast }.  \label{rhonondiagDef}
\end{eqnarray}
The density-matrix equation (DME) can be obtained by time differentiating of
$\rho _{mn}$ and employing Eq.\ (\ref{SchrEq}) that yields
\begin{eqnarray}
&&\hbar \dot{\rho}_{-1,-1}=\frac{i\Delta }{2}\left( \rho _{-1,1}-\rho
_{1,-1}\right) =\Delta \func{Im}\rho _{1,-1}  \nonumber \\
&&\hbar \dot{\rho}_{1,-1}=-iW(t)\rho _{1,-1}+\frac{i\Delta }{2}\left( \rho
_{1,1}-\rho _{-1,-1}\right)  \nonumber \\
&&{}+2iJ\sum_{k=0}^{N-1}\frac{(N-1)!}{(N-1-k)!k!}\left( N-2k-1\right)
C_{k}C_{k+1}^{\ast }.  \label{DME}
\end{eqnarray}
In the absence of interaction, $J=0,$ the last term in $\dot{\rho}_{-1,1}$
disappears and one obtains a DME for one isolated particle that is
equivalent to the one-particle Schr\"{o}dinger equation. Note that solving
this one-particle equation in the fast-sweep limit requires $\varepsilon \ll
1,$ in contrast to Eq.\ (\ref{SchrEq}) with $J=0$ that requires $\varepsilon
\ll 1/N.$

For $J\neq 0$ Eqs.\ (\ref{DME}) do not form a closed system of equations. In
this case Eqs.\ (\ref{DME}) can only be useful if an approximation for the
interaction term be found. In particular, one can consider the interaction
term perturbatively in $J$ and use
\begin{equation}
\rho _{mn}=\rho _{mn}^{(0)}+\delta \rho _{mn}  \label{rhoExpansion}
\end{equation}
where $\rho _{mn}^{(0)}$ is the density matrix without interaction that
satisfies
\begin{eqnarray}
&&\hbar \dot{\rho}_{-1,-1}^{(0)}=\Delta \func{Im}\rho _{1,-1}^{(0)}
\nonumber \\
&&\hbar \dot{\rho}_{1,-1}^{(\text{0)}}=-iW(t)\rho _{1,-1}^{(0)}+i\Delta
\left( \frac{1}{2}-\rho _{-1,-1}^{(0)}\right)  \label{DME0}
\end{eqnarray}
and $\delta \rho _{mn}$ is the correction term. The latter satisfies
equations similar to Eqs.\ (\ref{DME}) in which, however, $%
C_{k}C_{k+1}^{\ast }$ is replaced by its noninteracting-particle expression
following from Eq.\ (\ref{CFactorized}),
\begin{eqnarray}
&&C_{k}C_{k+1}^{\ast }=a_{-1}^{N-k}a_{1}^{k}\left( a_{-1}^{\ast }\right)
^{N-k-1}\left( a_{1}^{\ast }\right) ^{k+1}  \nonumber \\
&&{}\qquad =|a_{-1}|^{2(N-k-1)}|a_{1}|^{2k}a_{-1}a_{1}^{\ast }  \nonumber \\
&&\qquad =\left( \rho _{-1,-1}^{(0)}\right) ^{N-k-1}\left( 1-\rho
_{-1,-1}^{(0)}\right) ^{k}\rho _{-1,1}^{(0)}.  \label{CkCkp1}
\end{eqnarray}
With the use of the latter, the sum over $k$ in Eqs.\ (\ref{DME}) can be
performed to yield
\begin{eqnarray}
&&\hbar \delta \dot{\rho}_{-1,-1}=\Delta \func{Im}\delta \rho _{1,-1}
\nonumber \\
&&\hbar \delta \dot{\rho}_{1,-1}=-iW(t)\delta \rho _{-1,1}-i\Delta \delta
\rho _{-1,-1}  \nonumber \\
&&{}\qquad -2iJ_{0}\left( 1-2\rho _{-1,-1}^{(0)}\right) \rho _{1,-1}^{(0)}.
\label{DMEdelta}
\end{eqnarray}
Solving Eqs.\ (\ref{DME0}) and (\ref{DMEdelta}) perturbatively for $%
\varepsilon \ll 1$ results in Eq.\ (\ref{PNSmallJfast}).

In accord with the remark at the beginning of this section, $J$-perturbative
Eqs.\ (\ref{DME0}) and (\ref{DMEdelta}) can be solved in terms of the
hypergeometric functions for any $\varepsilon $ (For the standard LZS
problem, $J=0,$ this was done by Zener. \cite{zen32}) It can be shown that
in the slow-sweep limit this solution simplifies to
\begin{equation}
\left. \frac{dP}{d\rho }\right| _{\rho =0}=c\varepsilon e^{-\varepsilon
},\qquad \varepsilon \gg 1,\qquad \rho \equiv \frac{2J_{0}}{\Delta },
\label{dPdrhoslowsweep}
\end{equation}
where $c$ is a number and the parameter $\rho $ should not be confused with
the density matrix.

%TCIMACRO{
%\TeXButton{Fig-lzn-Pvsrho}{\begin{figure}[t]
%\unitlength1cm
%\begin{picture}(11,6)
%\centerline{\psfig{file=Fig-lzn-Pvsrho.eps,angle=-90,width=9cm}}
%\end{picture}
%\caption{ \label{Fig-lzn-Pvsrho}
%Numerically computed staying probability $P_{N}$ vs $\rho\equiv 2J_0/\Delta $ for $\varepsilon=1$ and different system sizes $N$, including the mean-field result for $N\to\infty$.}
%\end{figure}%
%}}%
%BeginExpansion
\begin{figure}[t]
\unitlength1cm
\begin{picture}(11,6)
\centerline{\psfig{file=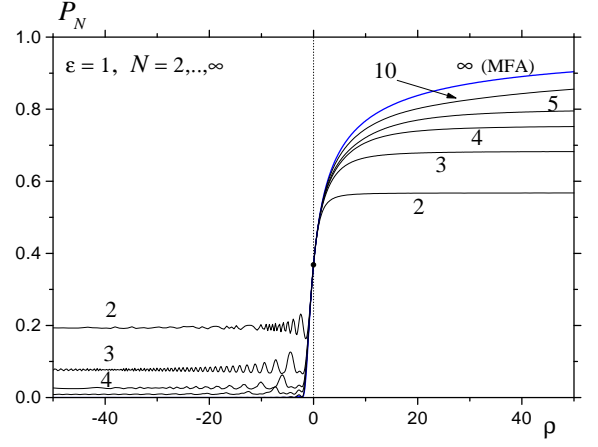,angle=-90,width=9cm}}
\end{picture}
\caption{ \label{Fig-lzn-Pvsrho}
Numerically computed staying probability $P_{N}$ vs $\rho\equiv 2J_0/\Delta $ for $\varepsilon=1$ and different system sizes $N$, including the mean-field result for $N\to\infty$.}
\end{figure}%
%
%EndExpansion

\section{The mean field approximation}

\label{Sec-mfa}

If each of tunneling particles interacts with all the other particles with a
coupling of the same sign, as is the case in our model for $N\gg 1,$ then
the mean-field approximation (MFA) should work well. The MFA employs a
one-particle density-matrix equation or a one-particle Schr\"{o}dinger
equation in which the interaction enters via the molecular field. For the
model of Eq.\ (\ref{Ham}) one has $W(t)\Rightarrow W_{\mathrm{eff}}(t)$
where
\begin{equation}
W_{\mathrm{eff}}(t)=W(t)+2J_{0}(1-2\rho _{-1,-1})  \label{WeffDef}
\end{equation}
in the DME of Eq.\ (\ref{DME0}), i.e.,
\begin{eqnarray}
&&\hbar \dot{\rho}_{-1,-1}=\Delta \func{Im}\rho _{1,-1}  \nonumber \\
&&\hbar \dot{\rho}_{1,-1}=-iW_{\mathrm{eff}}(t)\rho _{1,-1}+i\Delta \left(
\frac{1}{2}-\rho _{-1,-1}\right) .  \label{DMEMFA}
\end{eqnarray}
This is equivalent to the nonlinear Schr\"{o}dinger equation
\begin{eqnarray}
i\hbar \dot{C}_{-1} &=&\frac{1}{2}W_{\mathrm{eff}}(t)C_{-1}-\frac{\Delta }{2}%
C_{1}  \nonumber \\
i\hbar \dot{C}_{1} &=&-\frac{1}{2}W_{\mathrm{eff}}(t)C_{1}-\frac{\Delta }{2}%
C_{-1}  \label{NonlSchEq}
\end{eqnarray}
with $W_{\mathrm{eff}}(t)=W(t)+2J_{0}(1-2|C_{-1}|^{2}).$ It is interesting
to note that if one solves Eq.\ (\ref{DMEMFA}) perturbatively in $J$ \ using
Eq.\ (\ref{rhoExpansion}) one obtains Eqs.\ (\ref{DME0}) and (\ref{DMEdelta}%
). This implies that the MFA works well in the weak-coupling limit in our
model (see Fig.\ \ref{Fig-lzn-Pvsrho}).

In the case $N\gg 1$ one can assume that the state of the system is
described by a narrow packet in the $k$-space, as is indeed the case for
noninteracting particles, see Eq.\ (\ref{PNNoninteracting}). Then in Eq.\ (%
\ref{DME}) one can replace $\left( N-2k-1\right) \Rightarrow \left(
N-2k_{\max }-1\right) $ \ with $k_{\max }$ of Eq.\ (\ref{kmaxDef}), after
which the sum over $k$ converts to $\rho _{-1,1}$ according to the
definition in Eq.\ (\ref{rhonondiagDef}). This leads to Eq.\ (\ref{DMEMFA}).
One should, however, realize that this is no more than a heuristic
derivation since it was not proved that the state of a system \emph{with
interaction} is described by a narrow packet for $N\gg 1.$ On the other
hand, the narrow-packet assumption is quite plausible since the limit $N\gg
1 $ corresponds to the quasiclassical limit $S\gg 1$ of the equivalent model
of Eq.\ (\ref{HamS}) and the states of quasiclassical systems should be
narrow packets. It is remarkable that the density-matrix equation within the
MFA, Eq.\ (\ref{DMEMFA}), can be interpreted as a purely classical equation
of motion for a spin vector, the Landau-Lifschitz equation (LLE). Indeed,
rewriting Eq.\ (\ref{rhonondiagDef}) in terms of $c_{k}$ with the help of
Eq.\ (\ref{ckDef}) and adding Eq.\ (\ref{PtMapping}) one obtains the mapping
relations
\begin{eqnarray}
&&\rho _{1,-1}=\frac{1}{N}\sum_{k=0}^{N}l_{k,k+1}c_{k+1}^{\ast }c_{k}=\frac{%
\langle S_{+}\rangle }{2S}  \nonumber \\
&&\rho _{-1,-1}=\frac{1}{2}\left( 1-\frac{\langle S_{z}\rangle }{S}\right) .
\label{rhoSMapping}
\end{eqnarray}
Now one can see that the DME in the MFA, Eq.\ (\ref{DMEMFA}) is equivalent
to the LLE for the classical spin vector components
\begin{equation}
s_{z}=\frac{\langle S_{z}\rangle }{S},\qquad s_{+}=\frac{\langle
S_{+}\rangle }{S}  \label{sSavrIdent}
\end{equation}
that reads
\begin{eqnarray}
&&\hbar \dot{s}_{z}=-\Delta s_{y}  \nonumber \\
&&\hbar \dot{s}_{+}=-i\left[ W(t)+4J(S-1/2)s_{z}\right] s_{+}+i\Delta s_{z}.
\label{LLEzp}
\end{eqnarray}
The vector form of this LLE is
\begin{equation}
\mathbf{\dot{s}}=\gamma \left[ \mathbf{s\times H}_{\mathrm{eff}}\right]
,\qquad \mathbf{H}_{\mathrm{eff}}=-\frac{\partial \mathcal{H}_{\mathrm{eff}}%
}{\partial \mathbf{s}},  \label{LLE}
\end{equation}
where $\gamma =1/\hbar $ and the effective classical energy is given by
\begin{equation}
\mathcal{H}_{\mathrm{eff}}=-H_{z}(t)s_{z}-H_{x}s_{x}-D_{\mathrm{cl}}s_{z}^{2}
\label{HamClass}
\end{equation}
with
\begin{eqnarray}
&&H_{z}(t)=W(t),\qquad H_{x}=\Delta ,\qquad  \nonumber \\
&&D_{\mathrm{cl}}=D(S-1/2)=J(N-1)\equiv J_{0}  \label{ParIdentClass}
\end{eqnarray}
[cf. Eqs.\ (\ref{HamS}) and (\ref{ParIdent})]. The one-particle staying
probability $P_{N}$ is given by the formula
\begin{equation}
P_{N}(t)=\frac{1}{2}\left[ 1-s_{z}(t)\right]  \label{PtMappingClass}
\end{equation}
that is analogous to Eq.\ (\ref{PtMapping}).

%TCIMACRO{
%\TeXButton{Fig-lzn-PMFA-AF}{\begin{figure}[t]
%\unitlength1cm
%\begin{picture}(11,6)
%\centerline{\psfig{file=Fig-lzn-PMFA-AF.eps,angle=-90,width=9cm}}
%\end{picture}
%\caption{ \label{Fig-lzn-PMFA-AF}
%The mean-field solution for the one-particle staying probability $P_N$ for the AFMF coupling, $J_0\equiv (N-1)J<0$.
%}
%\end{figure}%
%}}%
%BeginExpansion
\begin{figure}[t]
\unitlength1cm
\begin{picture}(11,6)
\centerline{\psfig{file=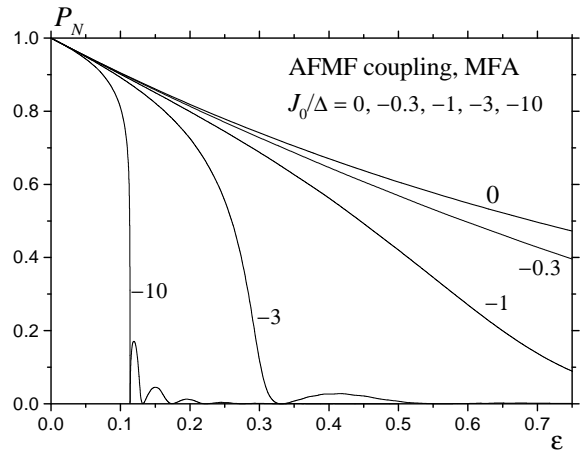,angle=-90,width=9cm}}
\end{picture}
\caption{ \label{Fig-lzn-PMFA-AF}
The mean-field solution for the one-particle staying probability $P_N$ for the AFMF coupling, $J_0\equiv (N-1)J<0$.
}
\end{figure}%
%
%EndExpansion

%TCIMACRO{
%\TeXButton{Fig-lzn-PWeffvst-AF-MFA-j10m}{\begin{figure}[t]
%\unitlength1cm
%\begin{picture}(11,6)
%\centerline{\psfig{file=Fig-lzn-Pvst-AF-MFA-j10m.eps,angle=-90,width=9cm}}
%\end{picture}
%\begin{picture}(11,6)
%\centerline{\psfig{file=Fig-lzn-Weffvst-AF-MFA-j10m.eps,angle=-90,width=9cm}}
%\end{picture}
%\caption{ \label{Fig-lzn-PWeffvst-AF-MFA-j10m}
%$a$ -- Time dependence of the staying probability $P_N$ for the AFMF coupling
%$J_0\equiv (N-1)J=-10$ for the first three complete conversion cases,
%$P_N(\infty)=0$ (see Fig.\ \protect\ref{Fig-lzn-PMFA-AF}).
%$b$ -- Same for the effective sweep $W_{\rm eff}$ of Eq.\ (\protect\ref{WeffDef}).
%}
%\end{figure}%
%}}%
%BeginExpansion
\begin{figure}[t]
\unitlength1cm
\begin{picture}(11,6)
\centerline{\psfig{file=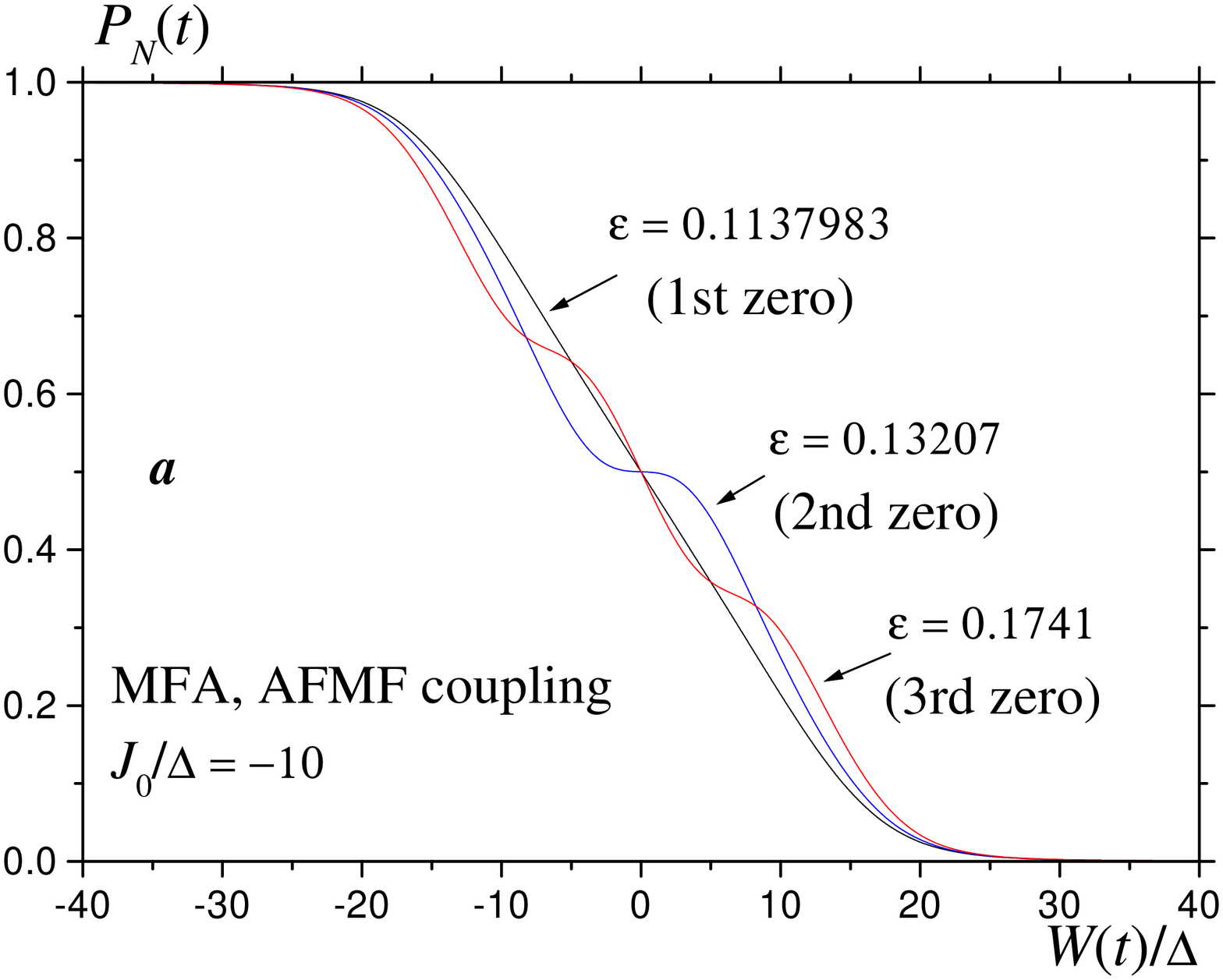,angle=-90,width=9cm}}
\end{picture}
\begin{picture}(11,6)
\centerline{\psfig{file=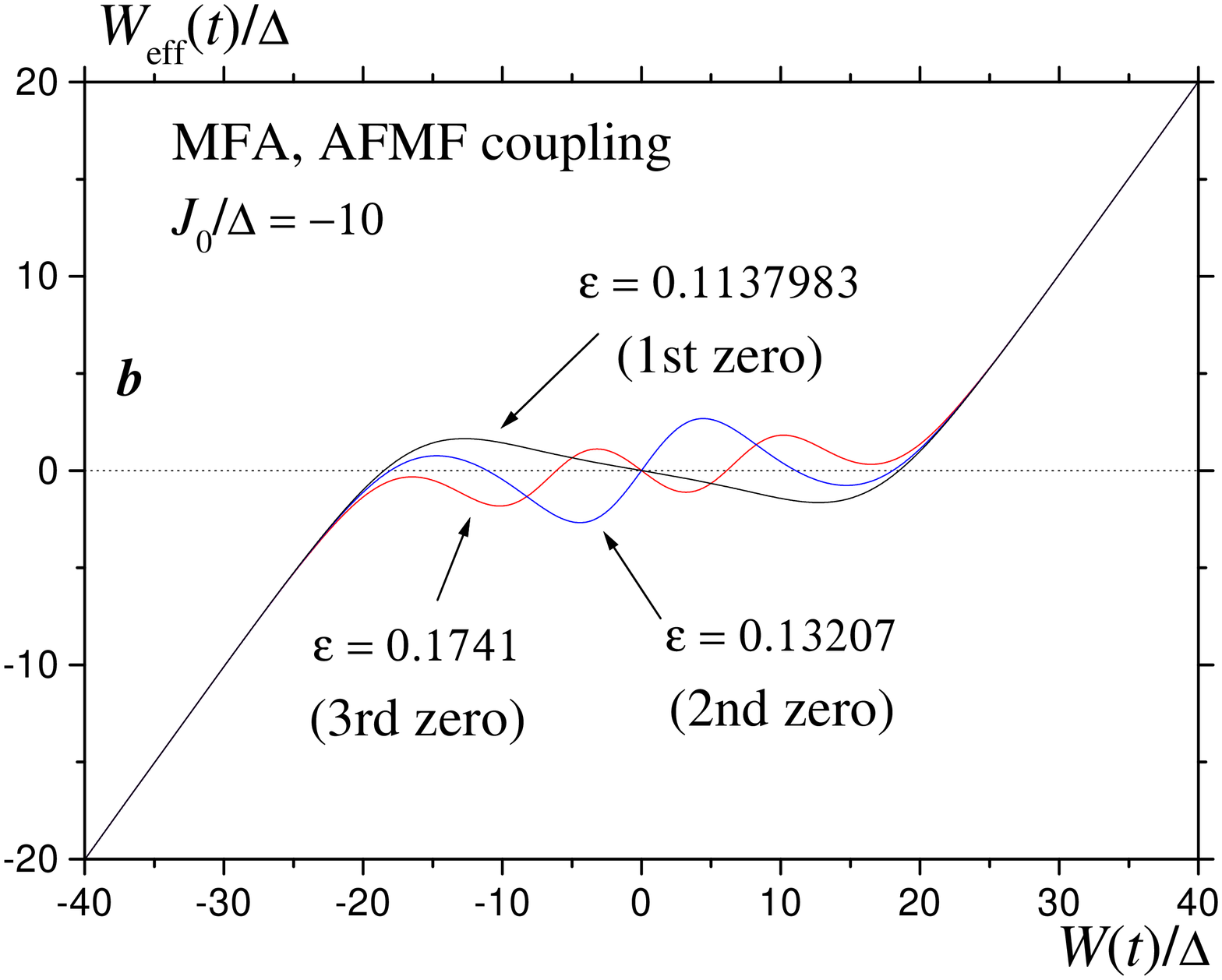,angle=-90,width=9cm}}
\end{picture}
\caption{ \label{Fig-lzn-PWeffvst-AF-MFA-j10m}
$a$ -- Time dependence of the staying probability $P_N$ for the AFMF coupling
$J_0\equiv (N-1)J=-10$ for the first three complete conversion cases,
$P_N(\infty)=0$ (see Fig.\ \protect\ref{Fig-lzn-PMFA-AF}).
$b$ -- Same for the effective sweep $W_{\rm eff}$ of Eq.\ (\protect\ref{WeffDef}).
}
\end{figure}%
%
%EndExpansion

Let us now analyze the mean-field solution of the LZS problem based on Eqs.\
(\ref{DMEMFA}) or (\ref{LLE}) and compare it with the exact solution of Eq.\
(\ref{SchrEqc}). We start with the AFMF coupling, for convenience. Fig.\ \ref
{Fig-lzn-PMFA-AF} shows the dependence of the one-particle staying
probability $P_{N}$ on the sweep-rate parameter $\varepsilon $ for different
coupling strengths $J_{0}.$ The curves on the plot are qualitatively similar
to those in the case of well-separated resonances, Fig.\ \ref
{Fig-lzn-separatedresonances} and they show that AFMF interaction boosts
transitions, especially for slow sweep, $\varepsilon \gg 1.$ The difference
is that in the MFA only the combined parameter $J_{0}=(N-1)J$ enters,
whereas for well-separated resonances the results depend on $N$ only and not
on $J.$ Another difference is that for strong enough interaction the
mean-field solution for $P_{N}$ oscillates and vanishes at some values of $%
\varepsilon .$ Oscillations of this kind have been found in the solution of
the LZS problem with nonlinear sweep.\cite{ternak97,garsch02prb} As was
explained in Ref.\ \onlinecite{garsch02prb}, oscillations arise when the
sweep slows down in the vicinity of the resonance so that (in the extreme
case) the particle oscillates between the two quantum states and the
resulting staying probability depends on the phase of this oscillation at
the end of the stay near the resonance. This is also the case for the model
with AFMF interaction in the MFA.Tunneling of particles creates a molecular
field that changes in the direction opposite to the sweep so that the
resulting effective sweep$\ W_{\mathrm{eff}}(t)$ of Eq.\ (\ref{WeffDef}) can
even become nonmonotonic (in this case the analytical method of Ref.\ %
\onlinecite{hamraemiysai00} based on the linearization of the problem near
the resonance and introducing an effective sweep rate breaks down). The MFA
solution becomes time symmetric in the case of complete conversion, $%
P_{N}(\infty )=0$ that for $J_{0}/\Delta =-10$ is realized for $\varepsilon
=0.1137983$, 0.13207, 0.1741, 0.2233, etc. The corresponding time
dependences of $P_{N}(t)$ and $W_{\mathrm{eff}}(t)$ are shown in Fig.\ (\ref
{Fig-lzn-PWeffvst-AF-MFA-j10m}). Although they resemble the complete
conversion solutions found in Ref.\ \onlinecite{garsch02}, there are no
apparent analytical solutions for this model. The difficulty of the present
mean-field model is that $W_{\mathrm{eff}}(t)$ is not known from the
beginning but is a solution of a self-consistent nonlinear problem.

\begin{figure}[t]
\unitlength1cm
\begin{picture}(11,6)
\centerline{\psfig{file=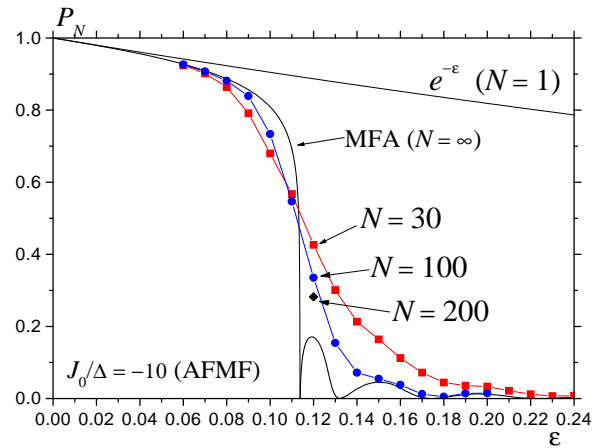,angle=-90,width=9cm}}
\end{picture}
\caption{ \label{Fig-lzn-P-AF-j10}
Comparison of the exact quantum and the MFA solutions for the one-particle staying probability $P_N$ for the AFMF coupling
showing partially slow convergence to the classical mean-field limit for $N\to\infty$.
}
\end{figure}

Comparison of the exact and the MFA solutions for the one-particle staying
probability $P_{N}$ for the antiferromagnetic coupling $J_{0}=-10$ is shown
in Fig.\ \ref{Fig-lzn-P-AF-j10}. One can see that the exact quantum solution
converges to the classical mean-field solution for $N\rightarrow \infty $,
although in the vicinity of the first complete-conversion point convergence
is extremely slow.

%TCIMACRO{
%\TeXButton{Fig-lzn-P-AF-j10}{\begin{figure}[t]
%\unitlength1cm
%\begin{picture}(11,6)
%\centerline{\psfig{file=Fig-lzn-P-AF-j10.eps,angle=-90,width=9cm}}
%\end{picture}
%\caption{ \label{Fig-lzn-P-AF-j10}
%Comparison of the exact quantum and the MFA solutions for the one-particle staying probability $P_N$ for the AFMF coupling
%showing partially slow convergence to the classical mean-field limit for $N\to\infty$.
%}
%\end{figure}%
%}}%
%BeginExpansion
%
%EndExpansion

%TCIMACRO{
%\TeXButton{Fig-lzn-PMFA-F}{\begin{figure}[t]
%\unitlength1cm
%\begin{picture}(11,6)
%\centerline{\psfig{file=Fig-lzn-PMFA-F.eps,angle=-90,width=9cm}}
%\end{picture}
%\caption{ \label{Fig-lzn-PMFA-F}
%The mean-field solution for the one-particle staying probability $P_N$ for the ferromagnetic coupling, $J_0\equiv (N-1)J>0$.
%The dashed line is the large-$\varepsilon$ asymptote of Eq.\ (\protect\ref{PN0alSmall}).
%}
%\end{figure}%
%}}%
%BeginExpansion
\begin{figure}[t]
\unitlength1cm
\begin{picture}(11,6)
\centerline{\psfig{file=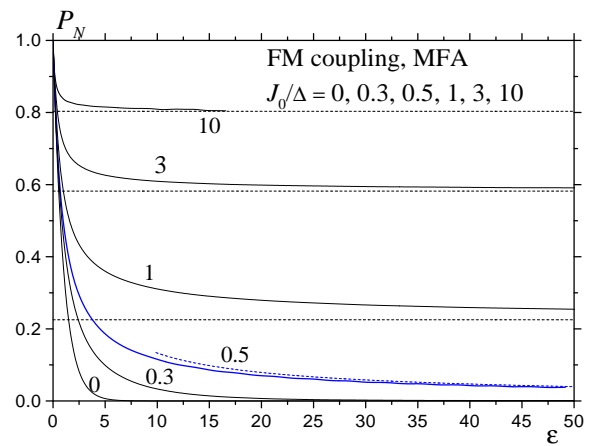,angle=-90,width=9cm}}
\end{picture}
\caption{ \label{Fig-lzn-PMFA-F}
The mean-field solution for the one-particle staying probability $P_N$ for the ferromagnetic coupling, $J_0\equiv (N-1)J>0$.
The dashed line is the large-$\varepsilon$ asymptote of Eq.\ (\protect\ref{PN0alSmall}).
}
\end{figure}%
%
%EndExpansion

%TCIMACRO{
%\TeXButton{Fig-lzn-P-F-j3}{\begin{figure}[t]
%\unitlength1cm
%\begin{picture}(11,6)
%\centerline{\psfig{file=Fig-lzn-P-F-j3.eps,angle=-90,width=9cm}}
%\end{picture}
%\caption{ \label{Fig-lzn-P-F-j3}
%Comparison of the exact quantum and the MFA solutions for the one-particle staying probability $P_N$ for the ferromagnetic coupling.
%}
%\end{figure}%
%}}%
%BeginExpansion
\begin{figure}[t]
\unitlength1cm
\begin{picture}(11,6)
\centerline{\psfig{file=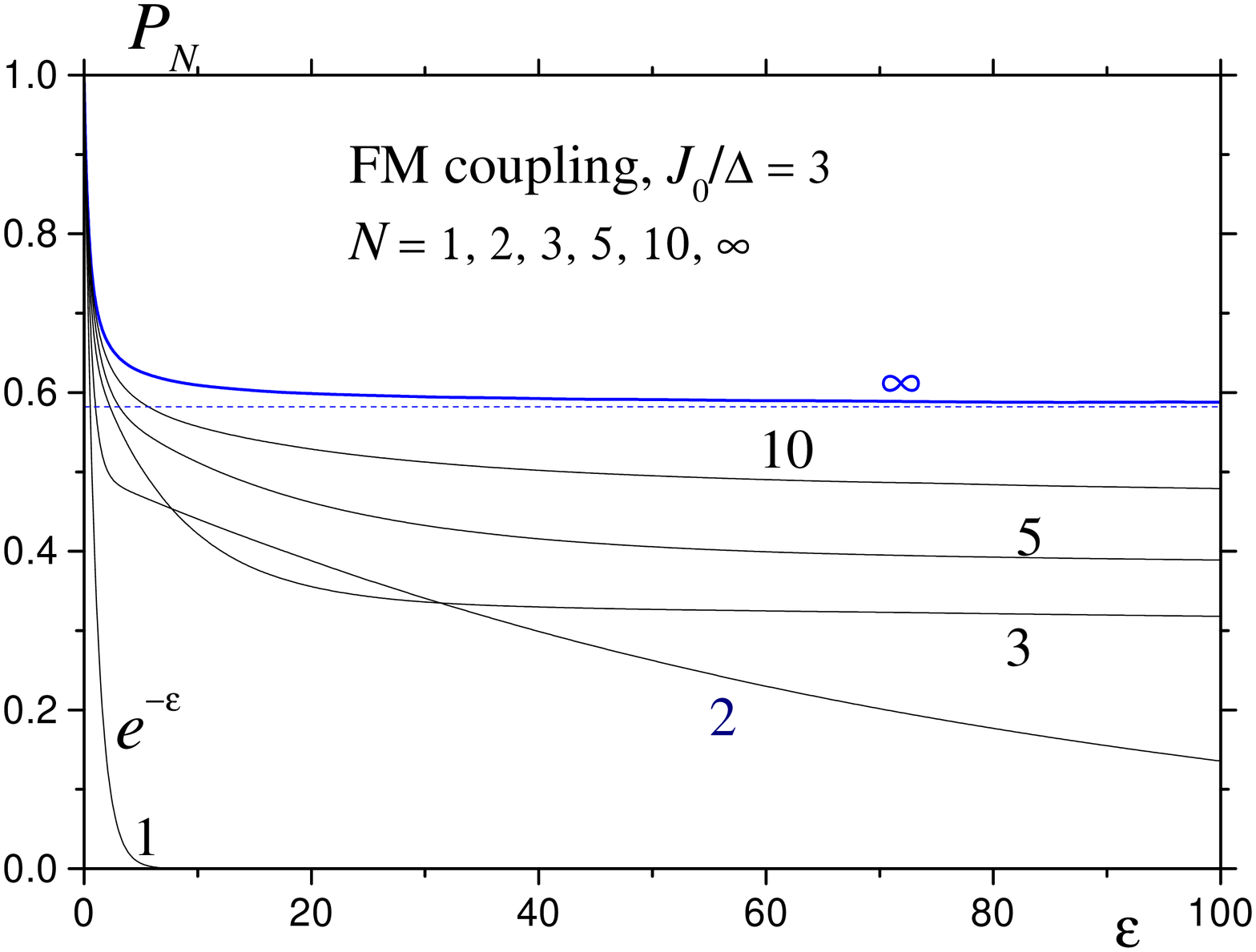,angle=-90,width=9cm}}
\end{picture}
\caption{ \label{Fig-lzn-P-F-j3}
Comparison of the exact quantum and the MFA solutions for the one-particle staying probability $P_N$ for the ferromagnetic coupling.
}
\end{figure}%
%
%EndExpansion

Let us now turn to the ferromagnetic interaction. The mean-field solution
for different interaction strengths is shown in Fig.\ \ref{Fig-lzn-PMFA-F}.
Again, qualitatively it is similar to the solution in the limit of
well-separated resonances, Eq.\ (\ref{PNF}) that is shown in Fig.\ \ref
{Fig-lzn-separatedresonances}. The difference is the same as for the AFMF
coupling: The MFA solution depends on $J_{0}\equiv (N-1)J$ whereas Eq.\ (\ref
{PNF}) depends on $N$ only. Another difference is that in the slow-sweep
limit the mean-field curve tends to a constant (for $J_{0}/\Delta >1/2)$
whereas all quantum curves tend to zero. The latter are in fact combinations
of several LZS exponentials corresponding to transitions of different orders
shown in Fig.\ \ref{Fig-lzn-En}a by dotted arrows (see also Fig.\ \ref
{Fig-lzn-P3Ft}). One can see, in particular, that the curve $N=2$ consist of
two different exponentials, the curve $N=3$ consist of three different
exponentials, etc. With increasing of $N,$ however, the coupling $%
J=J_{0}/(N-1)$ decreases and transitions become not well separated. In this
case dependence $P_{N}$ of $\varepsilon $ becomes a long-tale curve.

\section{Slow sweep in the MFA}

\label{Sec-slow-MFA}

Here we analytically consider the slow-sweep limit within the mean-field
approximation, to get more insights into the mechanism that leads to
suppressing transitions in the case of FM interactions (see Fig.\ \ref
{Fig-lzn-PMFA-F}) as well as into that of facilitating transitions and
probabillity oscillations in the case of the antiferromagnetic frustrating
interactions (see Fig.\ \ref{Fig-lzn-PMFA-AF}).

\subsection{Adiabatic case: Basic equations}

\begin{figure}[t]
\unitlength1cm
\begin{picture}(11,4.5)
\centerline{\psfig{file=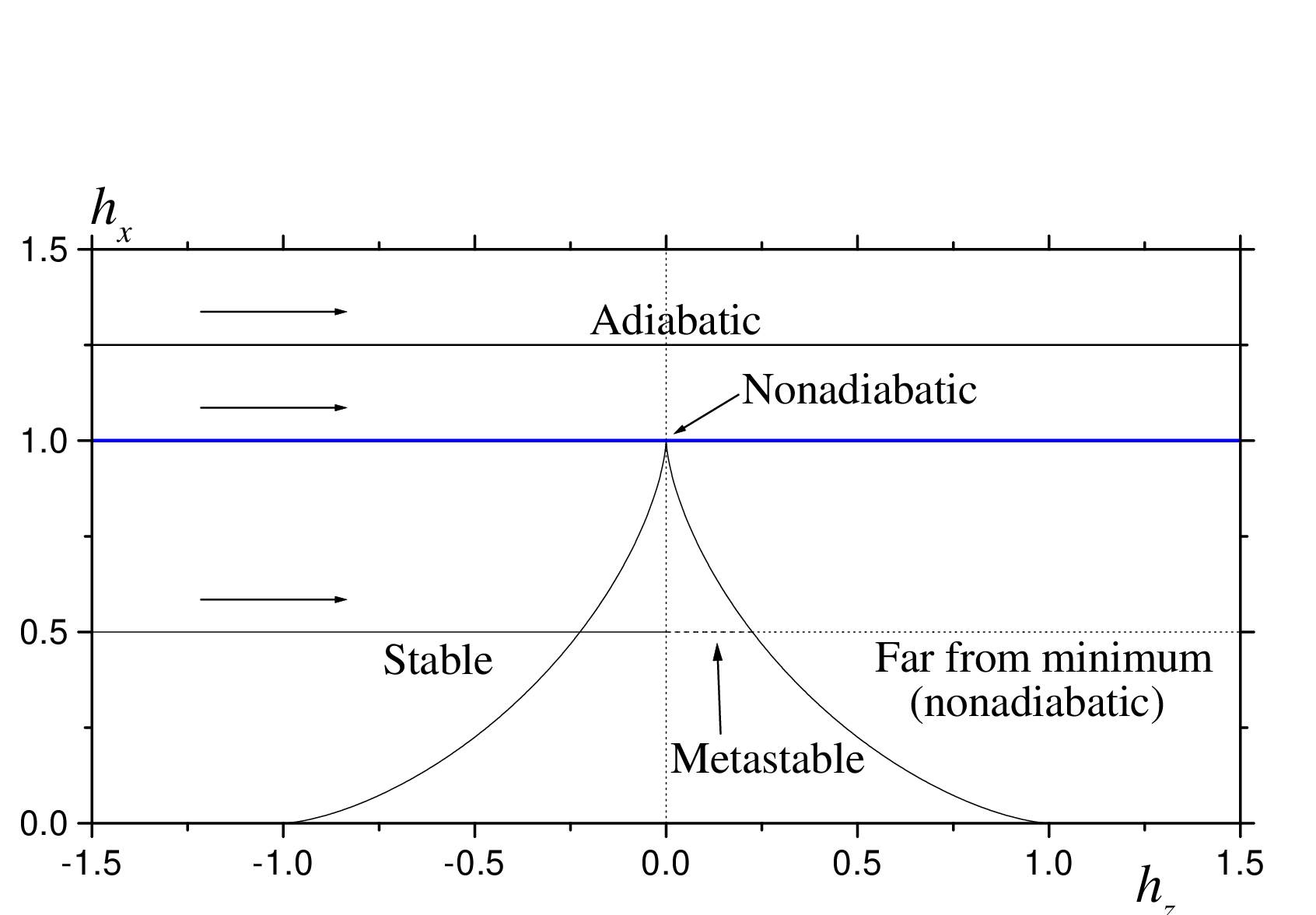,angle=-90,width=8cm}}
\end{picture}
\caption{ \label{Fig-lzn-P-F-Astroid}
Comparison of the exact quantum and the MFA solutions for the one-particle staying probability $P_N$ for the ferromagnetic coupling.
}
\end{figure}

\begin{figure}[t]
\unitlength1cm
\begin{picture}(11,6)
\centerline{\psfig{file=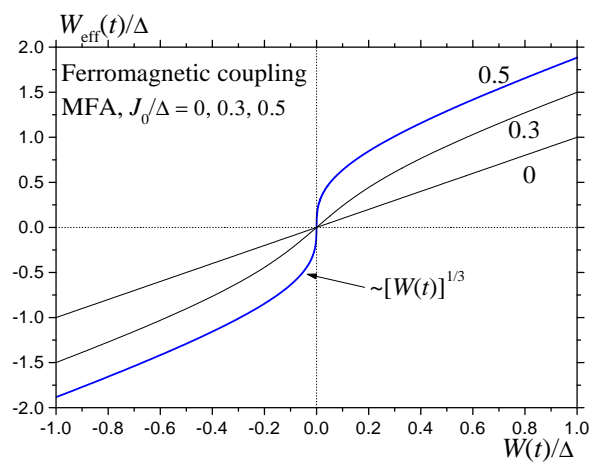,angle=-90,width=9cm}}
\end{picture}
\caption{ \label{Fig-lzn-P-F-Astroid-Weff}
Effective sweep in the adiabatic case for the ferromagnetic coupling of different strengths.
}
\end{figure}

For the one-particle LZS problem, a convenient method of treating the
slow-sweep limit is that using the adiabatic basis.\cite{garsch02prb} Its
advantage is that the probability to stay on the lowest adiabatic level is
at any time close to 1, so that one can make an appropriate approximation
after which the solution is obtained as a quadrature, including the case of
nonlinear sweep. Direct extention of this method for the present model
within the MFA is cumbersome, however, owing to the complicated
self-consistent nature of the adiabatic energy levels for a \emph{nonlinear}
Schr\"{o}dinger equation or \emph{nonlinear} density-matrix equation, Eq.\ (%
\ref{DME}). Fortunately, an alternative description based on the LLE, Eq.\ (%
\ref{LLE}), supports a physically transparent extension of the method. The
adiabatic basis for a Schr\"{o}dinger equation corresponds to the adiabatic
coordinate system for the classical problem of Eq.\ (\ref{LLE}). The $%
z^{\prime }$ axis of this coordinate system is oriented in the direction
minimizing the classical energy $\mathcal{H}_{\mathrm{eff}}$ of Eq.\ (\ref
{HamClass}) at any fixed time $t.$ It belongs to the $x$-$z$ plane and it
makes the angle $\theta (t)$ with the $z$ axis \ that satisfies the equation
\begin{equation}
h_{z}(t)\sin \theta +\nu \sin \theta \cos \theta -h_{x}\cos \theta =0
\label{EMinEq}
\end{equation}
following from $\partial \mathcal{H}_{\mathrm{eff}}/\partial \theta =0.$
Here $\nu =D_{\mathrm{cl}}/|D_{\mathrm{cl}}|=J/|J|$ and
\begin{equation}
h_{x}\equiv \frac{H_{x}}{2|D_{\mathrm{cl}}|}=\frac{\Delta }{2|J_{0}|},\qquad
h_{z}(t)\equiv \frac{H_{z}(t)}{2|D_{\mathrm{cl}}|}=\frac{W(t)}{2|J_{0}|}.
\label{hxhzDef}
\end{equation}
Eq.\ (\ref{EMinEq}) is well known in the theory of magnetism. In the AFMF
case $\nu <0$, its solution is unique. For the ferromagnetic interaction $%
\nu >0,$ Eq.\ (\ref{EMinEq}) has two solutions corresponding to the minimum
and to the maximum of $\mathcal{H}_{\mathrm{eff}}$ in the range of reduced
fields $h_{x}$ and $h_{z}$ outside the Stoner-Wohlfarth astroid \cite{stowoh}
(see Fig.\ \ref{Fig-lzn-P-F-Astroid})
\begin{equation}
h_{x}^{2/3}+h_{z}^{2/3}=1.  \label{Astroid}
\end{equation}
For $h_{x}$ and $h_{z}$ inside the astroid, Eq.\ (\ref{EMinEq}) has four
solutions corresponding to the stable and metastable minima as well as to
the saddle point and to the maximum of $\mathcal{H}_{\mathrm{eff}}.$ The
adiabatic solution corresponds to the minimum of the energy that changes as $%
h_{z}(t)$ is sweeped from $-\infty $ to $\infty $. For $h_{x}\geq 1$ the
system follows this adiabatic solution in the whole range of $h_{z}.$ In
this case dependence of the effective sweep $W_{\mathrm{eff}}$ of Eq.\ (\ref
{WeffDef}) or, here, $W_{\mathrm{eff}}=H_{z}+2D_{\mathrm{cl}}\cos \theta ,$
on the energy sweep $W$ is shown in Fig.\ \ref{Fig-lzn-P-F-Astroid-Weff}.
One can see that for $J_{0}/\Delta =0.5$ this dependence becomes
nonanalytic. This corresponds to the horizontal line in Fig.\ \ref
{Fig-lzn-P-F-Astroid} that touches the upper corner of the astroid. In
contrast, for $h_{x}<1$ the adiabatic solution exists only until the
crossing of the right branch of the astroid. After that the spin rotates
away from the disappeared metastable state and its behavior becomes
nonadiabatic. The latter case will be considered later while at first we
concentrate on the adiabatic situation.

%TCIMACRO{
%\TeXButton{Fig-lzn-P-F-Astroid}{\begin{figure}[t]
%\unitlength1cm
%\begin{picture}(11,4.5)
%\centerline{\psfig{file=Fig-lzn-P-F-Astroid.eps,angle=-90,width=8cm}}
%\end{picture}
%\caption{ \label{Fig-lzn-P-F-Astroid}
%Comparison of the exact quantum and the MFA solutions for the one-particle staying probability $P_N$ for the ferromagnetic coupling.
%}
%\end{figure}%
%}}%
%BeginExpansion
%
%EndExpansion

%TCIMACRO{
%\TeXButton{Fig-lzn-P-F-Astroid-Weff}{\begin{figure}[t]
%\unitlength1cm
%\begin{picture}(11,6)
%\centerline{\psfig{file=Fig-lzn-P-F-Astroid-Weff.eps,angle=-90,width=9cm}}
%\end{picture}
%\caption{ \label{Fig-lzn-P-F-Astroid-Weff}
%Effective sweep in the adiabatic case for the ferromagnetic coupling of different strengths.
%}
%\end{figure}%
%}}%
%BeginExpansion
%
%EndExpansion

The Landau-Lifshitz equation, Eq.\ (\ref{LLE}) can be rewritten in the
rotating adiabatic coordinate system as follows
\begin{equation}
\mathbf{\dot{s}}^{\prime }=\left[ \mathbf{s}^{\prime }\mathbf{\times }\left(
\gamma \mathbf{H}_{\mathrm{eff}}^{\prime }+\mathbf{\Omega }\right) \right]
,\qquad \mathbf{\Omega =}\dot{\theta}\mathbf{e}_{y},  \label{LLErotating}
\end{equation}
where $\dot{\theta}$ is time derivative of the appropriate solution of Eq.\ (%
\ref{EMinEq}). The reduced effective field
\begin{equation}
\mathbf{h}_{\mathrm{eff}}\equiv \frac{\mathbf{H}_{\mathrm{eff}}}{2|D_{%
\mathrm{cl}}|}=\left( h_{z}+\nu s_{z}\right) \mathbf{e}_{z}+h_{x}\mathbf{e}%
_{x},  \label{heff}
\end{equation}
in the adiabatic frame is calculated as follows
\begin{eqnarray}
\mathbf{h}_{\mathrm{eff}}^{\prime } &=&\left( h_{x}\cos \theta -h_{\mathrm{%
eff,}z}\sin \theta \right) \mathbf{e}_{x^{\prime }}  \nonumber \\
&&+\left( h_{x}\sin \theta +h_{\mathrm{eff,}z}\cos \theta \right) \mathbf{e}%
_{z^{\prime }}  \nonumber \\
h_{\mathrm{eff,}z} &=&h_{z}+\nu \left( -s_{x^{\prime }}\sin \theta
+s_{z^{\prime }}\cos \theta \right) .  \label{heffTrans}
\end{eqnarray}
With the help of Eq.\ (\ref{EMinEq}) components of $\mathbf{h}_{\mathrm{eff}%
}^{\prime }$ simplify to
\begin{eqnarray}
h_{\mathrm{eff,}x^{\prime }} &=&\left( 1-s_{z^{\prime }}+s_{x^{\prime }}\tan
\theta \right) \nu \sin \theta \cos \theta  \nonumber \\
h_{\mathrm{eff,}z^{\prime }} &=&h_{x}/\sin \theta -s_{x^{\prime }}\nu \sin
\theta \cos \theta  \nonumber \\
&&+\left( 1-s_{z^{\prime }}\right) \left( h_{z}-h_{x}\cot \theta \right)
\cos \theta .  \label{heffTransFinal}
\end{eqnarray}
Note that if the spin is directed along the $z^{\prime }$ axis, i.e., $%
s_{x^{\prime }}=s_{y}=0$ and s$_{z^{\prime }}=1,$ one has $h_{\mathrm{eff,}%
x^{\prime }}=0,$ that is consistent with the choice of the adiabatic
coordinate system. Introducing the dimensionless sweep variables
\begin{equation}
u\equiv \frac{W(t)}{\Delta }=\frac{vt}{\Delta }=\frac{h_{z}}{h_{x}},\qquad
\tilde{\varepsilon}\equiv \frac{\Delta ^{2}}{\hbar v}  \label{uDef}
\end{equation}
[$\tilde{\varepsilon}$ should not be confused with $\varepsilon =(\pi /2)%
\tilde{\varepsilon}$ of Eq.\ (\ref{PLZ})] one can rewrite Eq.\ (\ref
{LLErotating}) in the form
\begin{equation}
\frac{d\mathbf{s}^{\prime }}{du}=\left[ \mathbf{s}^{\prime }\mathbf{\times }%
\left( \tilde{\varepsilon}\frac{\mathbf{h}_{\mathrm{eff}}^{\prime }}{h_{x}}+%
\frac{d\theta }{du}\mathbf{e}_{y}\right) \right] .  \label{epstilDef}
\end{equation}

In the slow-sweep limit $\varepsilon \gg 1$ the solution of Eq.\ (\ref
{epstilDef}) is close to the adiabatic solution $s_{x^{\prime }}=s_{y}=0$
and s$_{z^{\prime }}=1.$ Hence one can linearize this equation near the
adiabatic solution:
\begin{eqnarray}
s_{z^{\prime }} &\Rightarrow &1,\qquad h_{\mathrm{eff,}z^{\prime
}}\Rightarrow h_{x}/\sin \theta  \nonumber \\
h_{\mathrm{eff,}x^{\prime }} &\Rightarrow &s_{x^{\prime }}\nu \sin ^{2}\theta
\label{Linearization}
\end{eqnarray}
that results in
\begin{eqnarray}
\frac{ds_{x^{\prime }}}{du} &=&\frac{\tilde{\varepsilon}}{\sin \theta }s_{y}-%
\frac{d\theta }{du}  \nonumber \\
\frac{ds_{y}}{du} &=&-\frac{\zeta \tilde{\varepsilon}}{\sin \theta }%
s_{x^{\prime }},  \label{LLElinearized}
\end{eqnarray}
where
\begin{equation}
\zeta \equiv 1-\rho \sin ^{3}\theta ,\qquad \rho \equiv 2J_{0}/\Delta .
\label{arhoDef}
\end{equation}
The factor $\zeta $ in the second of Eqs.\ (\ref{LLElinearized}) makes spin
precession elliptic for any nonzero coupling $J_{0},$ especially in the
extreme case $\rho =1$. This is an important difference from the model of
one tunneling particle with a nonlinear sweep, \cite{garsch02prb} where
precession remains always circular. It is convenient to introduce
\begin{equation}
\tilde{c}_{+}\equiv -\frac{1}{2}\left( \zeta ^{1/4}s_{x^{\prime }}+i\zeta
^{-1/4}s_{y}\right) ,\qquad \overline{\Omega }(u)\equiv \frac{\sqrt{\zeta }}{%
\sin \theta }  \label{cplusOmegaDef}
\end{equation}
and rewrite Eqs.\ (\ref{LLElinearized}) in the form
\begin{equation}
\frac{d\tilde{c}_{+}}{du}=-i\tilde{\varepsilon}\overline{\Omega }(u)\tilde{c}%
_{+}+\frac{1}{2}\frac{d\theta }{du}\zeta ^{1/4}.  \label{cplusEq}
\end{equation}
This differential equation can be easily solved with the initial condition $%
\tilde{c}_{+}(-\infty )=0,$ and the staying probability $P_{N}$ can be found
from Eq.\ (\ref{PtMappingClass}). Keeping in mind that at $t=\infty $ both
coordinate systems coincide whereby $\theta =0$ and $a=1$, one can rewrite $%
P_{N}\equiv P_{N}(\infty )$ within our linearized theory as
\begin{equation}
P_{N}\cong \frac{1}{4}\left[ s_{x^{\prime }}^{2}(\infty )+s_{y}^{2}(\infty )%
\right] =\left| \tilde{c}_{+}(\infty )\right| ^{2}.  \label{PNLinearized}
\end{equation}
The final general expression for $P_{N}$ reads
\begin{equation}
P_{N}\cong \left| \frac{1}{2}\int_{-\infty }^{\infty }du\frac{d\theta }{du}%
\zeta ^{1/4}\exp \left[ i\tilde{\varepsilon}\Phi (u)\right] \right| ^{2},
\label{PNfinalgeneral}
\end{equation}
where
\begin{equation}
\Phi (u)\equiv \int_{0}^{u}du^{\prime }\,\overline{\Omega }(u^{\prime }).
\label{PhiDef}
\end{equation}

In the absence of interaction $J=0,$ our solution recovers that for the
standard LZS problem in the slow-sweep limit up to the prefactor in front of
a small exponential. Indeed, in this case the solution of Eq.\ (\ref{EMinEq}%
) is
\begin{equation}
\cos \theta =\frac{h_{z}}{\sqrt{h_{x}^{2}+h_{z}^{2}}}=\frac{u}{\sqrt{1+u^{2}}%
},  \label{CoSiJ0}
\end{equation}
in Eq.\ (\ref{arhoDef}) one has $\rho =0$ and $\zeta =1,$ hence
\begin{equation}
\overline{\Omega }(u)=\sqrt{1+u^{2}},\qquad \frac{d\theta }{du}=-\frac{1}{%
1+u^{2}}=-\frac{1}{\overline{\Omega }^{2}(u)}.  \label{OmegaDerthetaJ0}
\end{equation}
One can see that in this case Eq.\ (\ref{cplusEq}) coincides with second of
Eqs.\ (24) of Ref.\ \onlinecite{garsch02prb} and Eq.\ (\ref{PNfinalgeneral})
coincides with Eq.\ (26) of Ref.\ \onlinecite{garsch02prb}, with $w^{\prime
}(u)=1$ and $\tilde{c}_{-}\Rightarrow 1.$ The latter is exactly the
approximation proposed in Ref.\ \onlinecite{garsch02prb} that for the
standard LZS problem reproduces the well-known result of Eq.\ (\ref{PLZ})
for $\varepsilon \gg 1$, however with a wrong prefactor: $P\approx (\pi
/3)^{2}e^{-\varepsilon }.$ In Ref.\ \onlinecite{garsch02prb} we have shown
how to correct the prefactor by accurately calculating $\tilde{c}_{-}.$

\subsection{Adiabatic case: The results }

\begin{figure}[t]
\unitlength1cm
\begin{picture}(11,6)
\centerline{\psfig{file=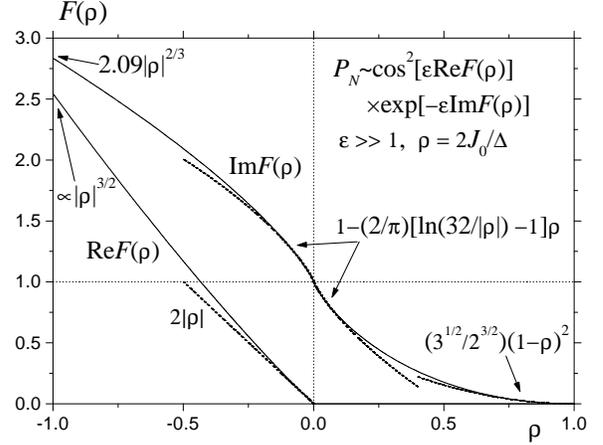,angle=-90,width=9cm}}
\end{picture}
\caption{ \label{Fig-lzn-Frho}
Function $F(\rho) $ of Eqs.\
(\protect\ref{PNExp}) and (\protect\ref{FrhoDef}).
}
\end{figure}

In the general case $J\neq 0$ Eq.\ (\ref{EMinEq}) does not yield an
analytical solution for $\theta (u).$ Fortunately, instead of $u$ one can
use $x\equiv \cos \theta $ as the integration variable on the interval $%
-1\leq x\leq 1$ and with the help of Eq.\ (\ref{EMinEq}) express $%
u=h_{z}/h_{x}$ as a function of $x.$ Even better is then to parametrize $x$
as $x=w/\sqrt{1+w^{2}},$ analogously to Eq.\ (\ref{CoSiJ0}). This brings
Eq.\ (\ref{PNfinalgeneral}) into the explicit form
\begin{equation}
P_{N}\cong \left| \frac{1}{2}\int_{-\infty }^{\infty }\frac{dw}{1+w^{2}}%
\left[ 1-\frac{\rho }{(1+w^{2})^{3/2}}\right] ^{1/4}e^{i\tilde{\varepsilon}%
\Phi (w)}\right| ^{2},  \label{PNx}
\end{equation}
where
\begin{equation}
\Phi (w)=\int_{0}^{w}dw\sqrt{1+w^{2}}\left[ 1-\frac{\rho }{(1+w^{2})^{3/2}}%
\right] ^{3/2}.  \label{Phix}
\end{equation}
In the absence of interaction, $\rho =0,$ one has $w=u$ and the formulas
above describe the standard LZS effect. For $\varepsilon \gg 1,$ the value
of $P_{N}$ is exponentially small and defined by the singularities of the
integrand in the complex plane of $w$
\begin{equation}
P_{N}=P_{N0}e^{-\varepsilon \func{Im}F(\rho )},  \label{PNExp}
\end{equation}
where
\begin{equation}
F(\rho )=\frac{4}{\pi }\int_{0}^{w_{c}}dw\sqrt{1+w^{2}}\left[ 1-\frac{\rho }{%
(1+w^{2})^{3/2}}\right] ^{3/2}.  \label{FrhoDef}
\end{equation}
For $\rho =0$ one obtains the LZS value $F(0)=i$. For the FM coupling the
relevant singularity is
\begin{equation}
w_{c}=i\sqrt{1-\rho ^{2/3}},\qquad 0<\rho <1,  \label{wcFM}
\end{equation}
that corresponds to vanishing of $\overline{\Omega }(u)$ of Eq.\ (\ref
{cplusOmegaDef}) due to that of the ellipticity factor $\zeta .$ For $\rho
>1,$ the motion of the spin becomes nonadiabatic (see Fig.\ \ref
{Fig-lzn-P-F-Astroid}), and this method does not apply any longer. For $%
0<\rho <1$ one has $\func{Re}F(\rho )=0$ and the limiting forms of Im$F(\rho
)$ are given by
\begin{equation}
\func{Im}F(\rho )\cong \left\{
\begin{array}{ll}
\displaystyle1-\frac{2}{\pi }\left( \ln \frac{32}{\rho }-1\right) \rho , &
\rho \ll 1 \\
\displaystyle\frac{\sqrt{3}}{2\sqrt{2}}(1-\rho )^{2}, & 1-\rho \ll 1.
\end{array}
\right.  \label{FrhoFMLims}
\end{equation}
For the AFMF coupling $\rho <0,$ there is a pair of relevant singularities
that also correspond to $\zeta =0$ and are given by
\begin{eqnarray}
w_{c\pm } &=&i\left( 1+|\rho |^{2/3}+|\rho |^{4/3}\right) ^{1/4}\exp \left(
\mp i\varphi _{c}\right)  \nonumber \\
\varphi _{c} &=&\frac{1}{2}\arctan \frac{\sqrt{3}|\rho |^{2/3}}{2+|\rho
|^{2/3}}.  \label{wcAFMF}
\end{eqnarray}
Note that $\func{Im}(w_{c\pm })>1$ for $\rho <0$ and thus these
singularities are \emph{further} from the real axis than the LZS singularity
at $w=i.$ The latter, however, makes no contribution since, as can be easily
checked, $\func{Im}F(\rho )=\infty $ for $w_{c}=i$ and $\rho <0.$ Analytical
calculation of the limiting forms of $F(\rho )$ is more cumbersome for the
antiferromagnetic coupling. For $|\rho |\ll 1$ one obtains
\begin{equation}
\func{Re}F(\rho )\cong 2|\rho |,\qquad \func{Im}F(\rho )\cong 1-\frac{2}{\pi
}\left( \ln \frac{32}{|\rho |}-1\right) \rho .  \label{FrhoAFMFsmall}
\end{equation}
Note that $\func{Im}F(\rho )$ is given by the same formula for $|\rho |\ll 1$
and both signs of $\rho .$ In the limit $|\rho |\gg 1$ the result has the
form
\begin{eqnarray}
F(\rho ) &\cong &\frac{2}{3\sqrt{\pi }}\frac{\Gamma (1/4)}{\Gamma (3/4)}%
\left( |\rho |^{3/2}-\frac{1}{4}|\rho |^{5/6}\right)  \nonumber \\
&&+\frac{\sqrt{\pi }\left( \sqrt{3}+i\right) }{\Gamma (5/3)\Gamma (11/6)}%
|\rho |^{2/3}  \nonumber \\
&\simeq &1.1|\rho |^{3/2}-0.3\,|\rho |^{5/6}+\left( 3.6+2.1i\right) |\rho
|^{2/3}.  \label{FrhoAFMFlarge}
\end{eqnarray}
The real and imaginary parts of $F(\rho )$ numerically calculated from Eq.\ (%
\ref{FrhoDef}), as well as the analytical limiting forms, are shown in Fig.\
\ref{Fig-lzn-Frho}.

%TCIMACRO{
%\TeXButton{Fig-lzn-Frho}{\begin{figure}[t]
%\unitlength1cm
%\begin{picture}(11,6)
%\centerline{\psfig{file=Fig-lzn-Frho.eps,angle=-90,width=9cm}}
%\end{picture}
%\caption{ \label{Fig-lzn-Frho}
%Function $F(\rho) $ of Eqs.\
%(\protect\ref{PNExp}) and (\protect\ref{FrhoDef}).
%}
%\end{figure}%
%}}%
%BeginExpansion
%
%EndExpansion

The prefactor $P_{N0}$ in Eq.\ (\ref{PNExp}) is determined for $\varepsilon
\gg 1$ by a close vicinity of the singularities $w_{c}.$ For the
ferromagnetic coupling the result is
\begin{equation}
P_{N0}\cong \frac{\pi ^{2}}{15\varepsilon \rho \sqrt{1-\rho ^{2/3}}},\qquad
0<\rho <1.  \label{P0F}
\end{equation}
For the AFMF coupling $\rho <0$ interference of the two contributions from $%
w_{c+}$ and $w_{c-}$ of Eq.\ (\ref{wcAFMF}) results in the oscillating
prefactor
\begin{equation}
P_{N0}\cong \frac{\pi ^{2}}{15\varepsilon |\rho ||w_{c}|}\cos ^{2}\left[
\frac{\varepsilon }{2}\func{Re}F(\rho )\right] ,\qquad \rho <0.  \label{P0AF}
\end{equation}
Oscillating prefactors of such a kind take place for\emph{\ one} tunneling
particle if the sweep is nonlinear and decelerating in the vicinity of the
resonance.\cite{garsch02prb} Here \emph{effective} nonlinearity of this type
occurs because of the negative coupling between tunneling particles, even
for the linear sweep.

Both Eqs.\ (\ref{P0F}) and (\ref{P0AF}) break down in the linit $\rho
\rightarrow 0$ since different singularities come close to each other. One
could work out the crossover from Eqs.\ (\ref{P0F}) and (\ref{P0AF}) to the
value $P_{N0}=(\pi /3)^{2}$ (see Ref.\ \onlinecite{garsch02prb}) at $\rho =0$
that takes place in a narrow region \TEXTsymbol{\vert}$\rho |\varepsilon
\sim 1.$ This makes no sence, however, since the result $P_{N0}=\pi /3$
differs from the exact LZS prefactor $P_{0}=1$ and it has to be improved by
taking into account nonlinear terms dropped during the derivation of Eq.\ (%
\ref{cplusEq}). The corresponding procedure for $\rho =0$ \ is described in
Ref.\ \onlinecite{garsch02prb}. In the present case $\rho \neq 0$ this would
be too involved and we don't try to do it. It is clear that the logarithmic
singularity of the exponent of Eq.\ (\ref{FrhoAFMFsmall}) should ve
compensated for by the singularity of the prefactor so that the staying
probability $P_{N}$ behaves linearly in $\rho $ near $\rho =0$ [see Eq.\ (%
\ref{dPdrhoslowsweep})].

\begin{figure}[t]
\unitlength1cm
\begin{picture}(11,6)
\centerline{\psfig{file=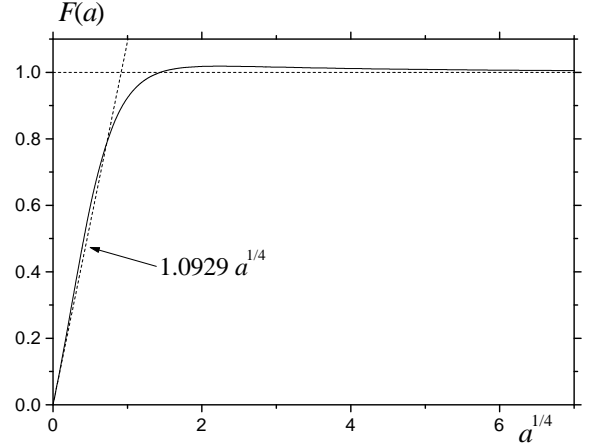,angle=-90,width=9cm}}
\end{picture}
\caption{ \label{Fig-lzn-FScaling-a}
Scaling function $f(a)$ of Eq.\ (\protect\ref{PN0Scaling})
}
\end{figure}

Let us now consider the case $1-\rho \ll 1$ in more detail since the
prefactor $P_{N0}$ of Eq.\ (\ref{P0F}) diverges at $\rho \rightarrow 1$.
Here one has $\func{Im}F(\rho )\ll 1,$ and the exponential decrease of $%
P_{N} $ is very slow. In this region according to Eq.\ (\ref{wcFM}) the
integral in Eq.\ (\ref{PNx}) is dominated by small $w$ and it can be
simplified to
\begin{eqnarray}
P_{N} &\cong &\frac{2}{3}\delta ^{3/2}\left| \frac{1}{2}\int_{-\infty
}^{\infty }dt\left( 1+t^{2}\right) ^{1/4}e^{ia\widetilde{\Phi }(t)}\right|
^{2}  \nonumber \\
\widetilde{\Phi }(t) &=&\int_{0}^{t}dt^{\prime }\left( 1+t^{\prime 2}\right)
^{3/2},  \label{PNt}
\end{eqnarray}
where
\begin{equation}
\delta \equiv 1-\rho \ll 1,\qquad t\equiv \sqrt{\frac{3}{2\delta }}w,\qquad
a\equiv \sqrt{\frac{2}{3}}\delta ^{2}\tilde{\varepsilon}.  \label{atDef}
\end{equation}
It is convenient to compute the integral in Eq.\ (\ref{PNt}) by shifting the
integration contour by $i$ to suppress oscillations of the integrand, i.e.,
to parametrize $t=i+z,$ $-\infty <z<\infty .$ With $\widetilde{\Phi }%
(i)=i3\pi /8$ this yields
\begin{equation}
P_{N}\cong P_{N0}e^{-a\func{Im}\widetilde{\Phi }(i)}=P_{N0}\exp \left( -%
\frac{3\pi }{8}a\right) ,  \label{PNdelta}
\end{equation}
where the exponent coincides with the previously obtained $[\sqrt{3}/(2\sqrt{%
2})]\delta ^{2}\varepsilon $ [see second line of Eq.\ (\ref{FrhoFMLims})].
The prefactor $P_{N0}$ can be calculated analytically for $a\gg 1$ and $a\ll
1.$ For $a\gg 1$ we need the small-$z$ expansions
\begin{eqnarray}
&&\delta \widetilde{\Phi }(z)\equiv \widetilde{\Phi }(i+z)-\widetilde{\Phi }%
(i)\cong \frac{4}{5}\left( -1+i\right) z^{5/2}  \nonumber \\
&&[1+(i+z)^{2}]^{1/4}\cong (-1)^{1/8}(2z)^{1/4}.  \label{SmallzExp}
\end{eqnarray}
Values of these functions for $z<0$ are obtained from $(-1)^{5/2}=-i$ and $%
(-1)^{1/4}=(1-i)/\sqrt{2}.$ After that calculation in Eq.\ (\ref{PNt})
yields
\begin{equation}
P_{N0}\cong \frac{2\pi }{15}\frac{\delta ^{3/2}}{a},\qquad a\gg 1
\label{PN0alarge}
\end{equation}
that is a limiting form of Eq.\ (\ref{P0F}). In the opposite limit $a\ll 1,$
the integral in Eq.\ (\ref{PNt}) is dominated by large $t,$ so that one can
use $\widetilde{\Phi }(t)\cong t^{4}/4$ and $\left( 1+t^{2}\right)
^{1/4}\cong t^{1/2}.$ This yields
\begin{equation}
P_{N0}\cong \cos ^{2}\left( \frac{3\pi }{16}\right) \frac{\Gamma ^{2}(3/8)}{6%
\sqrt{2}}\frac{\delta ^{3/2}}{a^{3/4}}\simeq \frac{0.747837}{\varepsilon
^{3/4}},\qquad \rho \rightarrow 1.  \label{PN0alSmall}
\end{equation}
It is convenient to represent $P_{N0}(a,\delta )$ in the whole range of $a$
with the help of the crossover function $f(a)$ according to
\begin{eqnarray}
&&P_{N0}(a,\delta )=\frac{2\pi }{15}\frac{\delta ^{3/2}}{a}f(a)  \nonumber \\
&&f(a)\cong \left\{
\begin{array}{ll}
1, & a\gg 1 \\
1.09294a^{1/4}, & a\ll 1.
\end{array}
\right.  \label{PN0Scaling}
\end{eqnarray}
Numerically computed function $f(a)$ is shown in Fig.\ (\ref
{Fig-lzn-FScaling-a}).

%TCIMACRO{
%\TeXButton{Fig-lzn-FScaling-a.eps}{\begin{figure}[t]
%\unitlength1cm
%\begin{picture}(11,6)
%\centerline{\psfig{file=Fig-lzn-FScaling-a.eps,angle=-90,width=9cm}}
%\end{picture}
%\caption{ \label{Fig-lzn-FScaling-a}
%Scaling function $f(a)$ of Eq.\ (\protect\ref{PN0Scaling})
%}
%\end{figure}%
%}}%
%BeginExpansion
%
%EndExpansion

%TCIMACRO{
%\TeXButton{Fig-lzn-Phx}{\begin{figure}[t]
%\unitlength1cm
%\begin{picture}(11,6)
%\centerline{\psfig{file=Fig-lzn-Phx.eps,angle=-90,width=9cm}}
%\end{picture}
%\caption{ \label{Fig-lzn-Phx}
%$P_N$ in the slow-sweep limit $\varepsilon\to\infty$ vs $h_x=1/\rho=\Delta/(2J_0)$.
%}
%\end{figure}%
%}}%
%BeginExpansion
\begin{figure}[t]
\unitlength1cm
\begin{picture}(11,6)
\centerline{\psfig{file=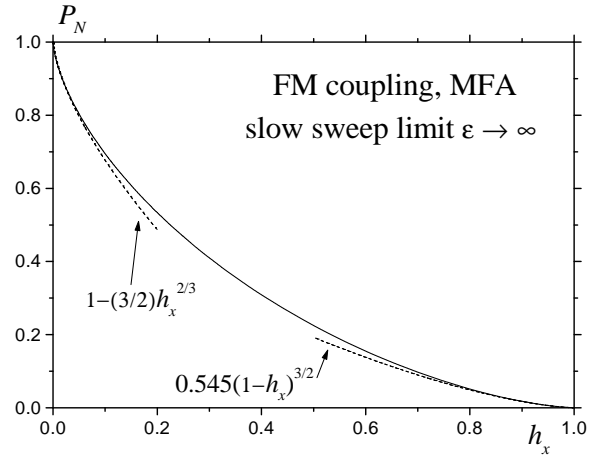,angle=-90,width=9cm}}
\end{picture}
\caption{ \label{Fig-lzn-Phx}
$P_N$ in the slow-sweep limit $\varepsilon\to\infty$ vs $h_x=1/\rho=\Delta/(2J_0)$.
}
\end{figure}%
%
%EndExpansion

\subsection{Strong ferromagnetic interactions}

For stronger ferromagnetic interactions, $\rho >1,$ the adiabatic
approximation breaks down since the spin approximately follows the initial
energy minimum that becomes unstable at some $h_{z}(t)>0$ (see Fig.\ \ref
{Fig-lzn-P-F-Astroid}) and than it performes a large motion that does not
approach the new stable energy minimum and thus cannot be linearized around
it. While the problem becomes much more complicated in this case, one can
still find analytically $\stackunder{\varepsilon \rightarrow \infty }{\lim }%
P_{N}$ that is nonzero$.$ To this end, one can represent the LLE for the
spin, Eq.\ (\ref{LLE}), in the Hamiltonian form in terms of the canonical
angle variables $\left\{ \cos \theta ,\varphi \right\} \Leftrightarrow
\left\{ p,q\right\} $ (in the arbitrary frame)
\begin{equation}
\frac{d}{dt}\cos \theta =-\frac{\partial \mathcal{H}}{\partial \varphi }%
,\qquad \frac{d}{dt}\varphi =\frac{\partial \mathcal{H}}{\partial \cos
\theta }  \label{HamEq}
\end{equation}
and use conservation of the action
\begin{equation}
\mathcal{S}=\oint pdq=\oint \cos \theta d\varphi  \label{ActionDef}
\end{equation}
over the period of motion for very slowly changing parameters of the system
[here $h_{z}(t)]$. Since in the final state $h_{z}\rightarrow \infty $ the
spin precesses around the $z$ axis with a constant value of $s_{z}(\infty
)=\cos \left[ \theta (\infty )\right] $ that is related to the staying
probability $P_{N}$ by Eq.\ (\ref{PtMappingClass}) and the corresponding
action is simply $\mathcal{S}(\infty )=2\pi \cos \theta (\infty ),$ one
obtains
\begin{equation}
P_{N}=\frac{1}{2}\left( 1-\frac{1}{2\pi }\mathcal{S}\right) ,\qquad
\varepsilon \rightarrow \infty ,  \label{PNActionDef}
\end{equation}
where $\mathcal{S}$ is the action over the trajectory that starts from the
metastable energy minimum that is on the verge of being unstable
\begin{equation}
s_{zs}=-(1-h_{x}^{2/3})^{1/2},\qquad s_{xs}=h_{x}^{1/3},\qquad s_{ys}=0.
\label{sStarting}
\end{equation}
This trajectory can be found from the conservation of energy,
\begin{equation}
-s_{z}^{2}-2(1-h_{x}^{2/3})^{3/2}s_{z}-2h_{x}s_{x}=1-3h_{x}^{2/3}.
\label{EnergyConservation}
\end{equation}
Analysis shows that in the range $0<h_{x}<3\sqrt{3}/8$ this trajectory
encircles the $z$ axis, thus the $z$ axis can be used as the polar axis ($%
s_{z}=\cos \theta ,$ $s_{x}=\sin \theta \cos \varphi ,$ $s_{y}=\sin \theta
\sin \varphi $) to compute $\mathcal{S}.$ In the overlapping range $%
1/8<h_{x}<1$ the trajectory encircles the $x$ axis that can be used as the
polar axis ($s_{x}=\cos \theta ,$ $s_{z}=-\sin \theta \cos \varphi ,$ $%
s_{y}=\sin \theta \sin \varphi $). With these choices, $\cos \theta $ can be
found numerically from Eq.\ (\ref{EnergyConservation}) as a function of $%
\varphi $ and the action $\mathcal{S}$ can be computed from Eq.\ (\ref
{ActionDef}). The result for $P_{N}$ vs $h_{x}$ is shown in Fig.\ \ref
{Fig-lzn-Phx} with asymptotes
\begin{equation}
P_{N}\cong \left\{
\begin{array}{ll}
1-(3/2)h_{x}^{2/3}, & h_{x}\ll 1 \\
0.544861\left( 1-h_{x}\right) ^{3/2}, & 1-h_{x}\ll 1.
\end{array}
\right.  \label{PhxLims}
\end{equation}

\section{Discussion}

The simplified model of interacting tunneling particles that was considered
above allows to quantify the influence of interaction on the
Landau-Zener-Stueckelberg staying probability $P$. It was done here by
numerically solving the problem for up to $N=200$ interacting particles as
well as using a number of analytical and half-analytical approaches,
including the mean-field limit $N\rightarrow \infty $.

In accord with physical expectations and considerations of the exact levels
of the system in Fig.\ \ref{Fig-lzn-En}a, the ferromagnetic coupling tends
to suppress LZS transitions to another bare energy level (i.e., the state on
the other side of the energy barrier), in agreement with Ref.\ %
\onlinecite{hamraemiysai00}. For $N\rightarrow \infty $ there is a critical
value of the coupling above which the staying probability $P$ does not go to
zero in the slow-sweep limit $\varepsilon \rightarrow \infty $, in contrast
with the standard LZS case. For finite $N$ the dependence of $P$ on $%
\varepsilon $ is a sum of many exponentials with greatly differing
relaxation rates that is very slow approaching zero (see Fig.\ \ref
{Fig-lzn-P-F-j3}). The same should be the case for more realistic
interactions of the ferromagnetic type.

The negative coupling in our model (that corresponds to the
antiferromagnetic frustrating coupling) tends to boost the LZS transitions.
In the limit $N\rightarrow \infty $ the staying probability $P$ even turns
to zero at \emph{finite} values of the sweep rate while the \emph{effective}
sweep rate becomes an odd function of time with retardation in the
resonance-crossing region (see Fig.\ \ref{Fig-lzn-PWeffvst-AF-MFA-j10m}). On
the other hand, models with more realistic antiferromagnetic interactions
such as the nearest-neighbor interaction have the energy-level scheme
strongly differing from that shown in Fig.\ \ref{Fig-lzn-En}b. Preliminary
results show that such interactions tend to hamper LZS transitions instead
of boosting it, although not to such an extent as ferromagnetic interactions.

Theoretically the model considered here is interesting since it maps on the
problem of a single large spin $S=N/2$ and thus the mean-field limit $%
N\rightarrow \infty $ corresponds to the classical limit $S\rightarrow
\infty $ for the large spin. It would be very interesting to study
deviations from the mean-field solutions for large but finite $N.$ These
deviations can be very large, as can be seen in Fig.\ \ref{Fig-lzn-P-AF-j10}
in the region $\varepsilon \simeq 0.11.$ How quantitatively well does the
MFA work for model systems with more complicated interactions and for
realistic systems remains unclear, and it is an interesting topic for
further work.

\section{\protect\bigskip Acknowledgments}

Many useful discussions with Rolf Schilling are greatfully acknowledged.

%\bigskip

%\bibliographystyle{prsty}
%\bibliography{gar-general,gar-tunneling,gar-own,gar-oldworks}

\begin{thebibliography}{99}
\bibitem{lan32}  {L. D. Landau}, Phys. Z. Sowjetunion \textbf{2}, 46 (1932).

\bibitem{zen32}  C. Zener, Proc. R. Soc. London A \textbf{137}, 696 (1932).

\bibitem{stu32}  E.~C.~G. Stueckelberg, Helv. Phys. Acta \textbf{5}, 369
(1932).

\bibitem{crohug77}  {D. S. F. Crothers and J. G Huges}, J. Phys. B \textbf{10%
}, L557 (1977).

\bibitem{werses99}  {W. Wernsdorfer and R. Sessoli}, Science \textbf{284},
133 (1999).

\bibitem{weretal00epl}  {W. Wernsdorfer, R. Sessoli, A. Caneshi, D.
Gatteschi, and A. Cornia}, Europhys. Lett. \textbf{50}, 552 (2000).

\bibitem{hamraemiysai00}  {A. Hams, H. De Raedt, S. Miyashita, and K. Saito}%
, Phys. Rev. B \textbf{62}, 13880 (2000).

\bibitem{garsch02prb}  {D. A. Garanin and R. Schilling}, Phys. Rev. B
\textbf{66}, 174438 (2002).

\bibitem{akusch92}  {V. M. Akulin and W. P. Schleich}, Phys. Rev. A \textbf{%
46}, 4110 (1992).

\bibitem{dobzve97}  {V. V. Dobrovitski and A. K. Zvezdin}, Europhys. Lett.
\textbf{38}, 377 (1997).

\bibitem{gar91jpa}  {D. A. Garanin}, J. Phys. A \textbf{24}, L61 (1991).

\bibitem{ternak97}  {Y. Teranishi and H. Nakamura}, J. Chem. Phys. \textbf{%
107}, 1904 (1997).

\bibitem{garsch02}  {D. A. Garanin and R. Schilling}, Europhys. Lett.
\textbf{59}, 7 (2002).

\bibitem{stowoh}  {E. C. Stoner and E. P. Wohlfarth}, Philos. Trans. R. Soc.
London, Ser. A \textbf{\ 240}, 599 (1948); IEEE Trans. Magn. \textbf{MAG-27}%
, 3475 (1991).
\end{thebibliography}

\end{document}